\documentclass[11pt,english]{article}
\usepackage[T1]{fontenc}
\usepackage[latin9]{luainputenc}
\usepackage{geometry}
\usepackage{array}
\usepackage{float}
\usepackage{multirow}
\usepackage{amsmath}
\usepackage{amsthm}
\usepackage{amssymb}
\usepackage{graphicx}
\usepackage{setspace}
\usepackage{esint}
\usepackage[authoryear]{natbib}
\makeatletter

\providecommand{\tabularnewline}{\\}
\floatstyle{ruled}
\newfloat{algorithm}{tbp}{loa}
\providecommand{\algorithmname}{Algorithm}
\floatname{algorithm}{\protect\algorithmname}

\theoremstyle{plain}
\newtheorem{thm}{\protect\theoremname}
  \theoremstyle{plain}
  \newtheorem{prop}[thm]{\protect\propositionname}

\usepackage{amsmath}

\usepackage{algorithm,algpseudocode}
\date{}

\@ifundefined{showcaptionsetup}{}{%
 \PassOptionsToPackage{caption=false}{subfig}}
\usepackage{subfig}
\makeatother

\usepackage{babel}
  \providecommand{\propositionname}{Proposition}
\providecommand{\theoremname}{Theorem}

\begin{document}

\title{A novel approach for fusion of heterogeneous sources of data}

\author{Mostafa Reisi Gahrooei$^{1}$, Hao Yan$^{2}$, Kamran Paynabar$^{1}$,
Jianjun Shi$^{1}$\\
$^{1}$Georgia Institute of Technology, Atlanta, GA 30332, USA\\
$^{2}$Arizona State University, Tempe, AZ 85281, USA}
\maketitle
\begin{abstract}
With advancements in sensor technology, heterogeneous sets of data
such as those containing scalars, waveform signals, images, and even
structured point clouds, are becoming increasingly popular. Developing
statistical models based on such heterogeneous sets of data to represent
the behavior of an underlying system can be used in the monitoring,
control, and optimization of the system. Unfortunately, available
methods only focus on the scalar and curve data and do not provide
a general framework for integrating different sources of data to construct
a model. This paper addresses the problem of estimating a process
output, measured by a scalar, curve, image, or structured point cloud
by a set of heterogeneous process variables such as scalar process
setting, sensor readings, and images. We introduce a general multiple
tensor-on-tensor regression (MTOT) approach in which each set of input
data (predictor) and output measurements are represented by tensors.
We formulate a linear regression model between the input and output
tensors and estimate the parameters by minimizing a least square loss
function. In order to avoid overfitting and reduce the number of parameters
to be estimated, we decompose the model parameters using several bases
that span the input and output spaces. Next, we learn the bases and
their spanning coefficients when minimizing the loss function using
a block coordinate descent algorithm combined with an alternating
least square (ALS) algorithm. We show that such a minimization has
a closed-form solution in each iteration and can be computed very
efficiently. Through several simulation and case studies, we evaluate
the performance of the proposed method. The results reveal the advantage
of the proposed method over some benchmarks in the literature in terms
of the mean square prediction error.
\end{abstract}

\section{Introduction}

With advancements in sensor technology, heterogeneous sets of data
containing scalars, waveform signals, images, etc. are more and more
available. For example, in semiconductor manufacturing, machine/process
settings (scalar variables), sensor readings in chambers (waveform
signals), and wafer shape measurements (images) may be collected to
model and monitor the process. In this article, we refer to non-scalar
variables as high-dimensional (HD) variables. Analysis and evaluation
of such heterogeneous data sets can benefit many applications, including
manufacturing processes \citep{szatvanyi2006multivariate,wojcik2009combustion},
food industries \citep{yu2003multivariate}, medical decision-making
\citep{bellon1995experimental}, and structural health monitoring
\citep{balageas2010structural}. Specifically, analysis of such data
may lead to the construction of a statistical model that estimates
a variable (including HD) based on several other variables (regression
model). Unfortunately, most works in regression modeling only consider
scalars and waveform signals \citep{liang2003relationship,Ramsay2005,fan2014functional,luo2017function},
without including images or structured point clouds. However, in several
applications, images or structured point clouds are available. For
example, materials scientists are interested in constructing a link
between process variables, e.g., the temperature and pressure under
which a material is operating, and the microstructure of the material
\citep{khosravani2017development}, which is often represented by
an image or a variation of the microstructure image obtained by two-point
statistics. Generating such a linkage model requires regression analysis
between scalar and waveform process variables as inputs and an image.
For more detailed information and a case study, please refer to \citep{Gorgannejad2018}.

\begin{figure}
\begin{centering}
\subfloat[\label{fig:waferShape-1}]{\includegraphics[width=0.4\columnwidth]{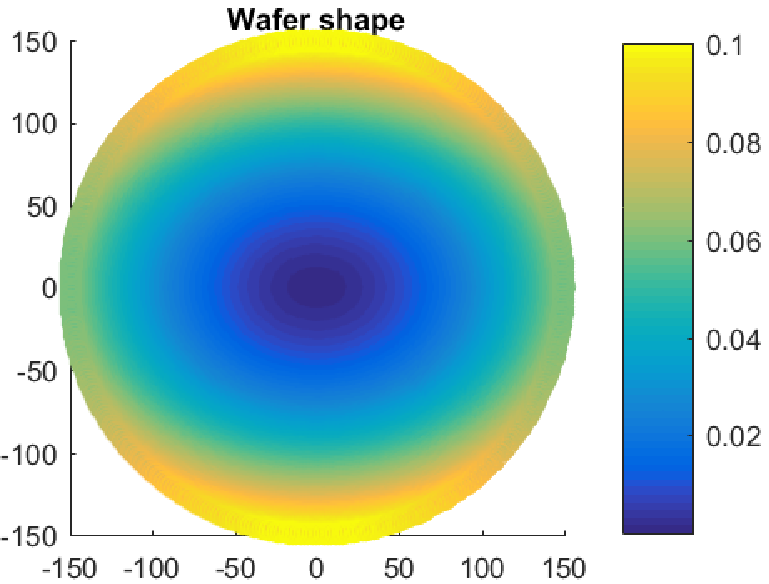}}\subfloat[\label{fig:overlayError-1}]{\includegraphics[width=0.4\columnwidth]{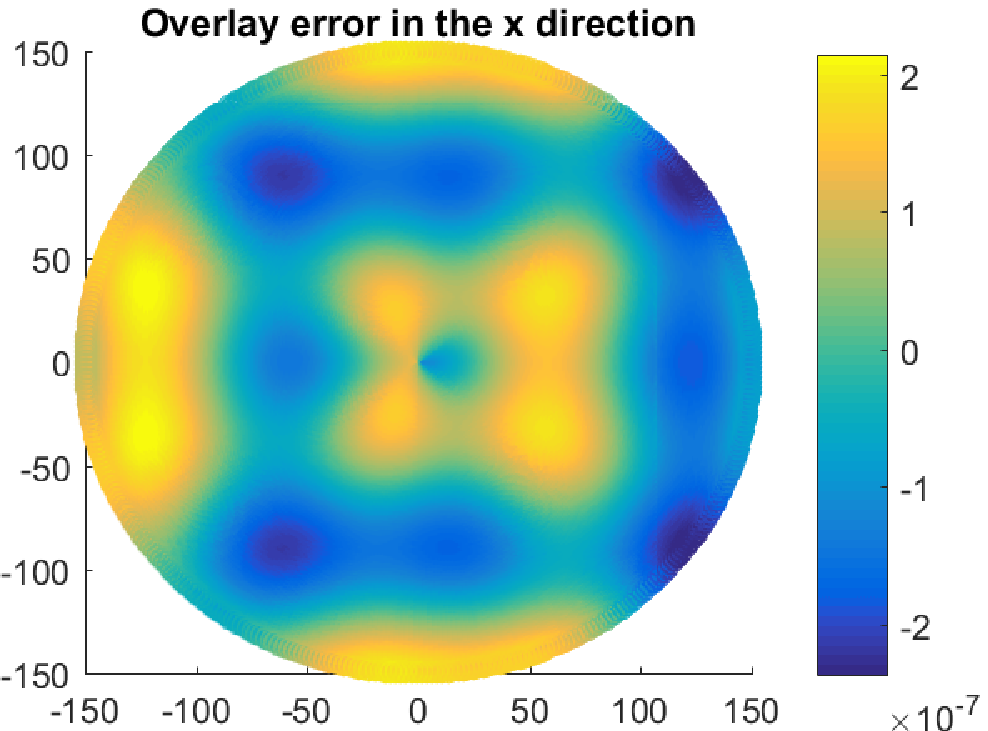}

}
\par\end{centering}
\caption{Examples of the (a) wafer shape and (b) x coordinate overlay error.
All measurements are in millimeters (mm). The wafer in (a) has an
overall bow shape, with several high-order wave-form patterns that
cannot be seen. Figure (b) represents the misalignment of two layers
in the x coordinate. The dark blue represents a large error in the
negative direction of the x coordinate, and light yellow represents
a large error in the positive direction of the x coordinate. \label{fig:Examples-of-wafer-shape-overlay-1}}
\end{figure}

As another example, in semiconductor manufacturing, overlay errors
(defined as the difference between in-plane displacements of two layers
of a wafer) are directly influenced by the shape of the wafer before
the lithographic process. In this process, both the wafer shape and
the overlay error (in the \textit{x} and \textit{y} directions) can
be represented as images as shown in Figure \ref{fig:Examples-of-wafer-shape-overlay-1}.
Prediction of the overlay error across a wafer based on the wafer
shape can be fed forward to exposure tools for specific corrections
\citep{turner2013role}. In order to predict the overlay error based
on the wafer shape deformation, an image-on-image statistical model
is required to capture the correlation between the wafer overlay and
shape.

In addition to the space and computational issues caused by the large
size of the HD variables, the challenge of developing an accurate
model for a process with a heterogeneous group of variables is twofold:
Integrating variables of different forms (e.g., scalars, images, curves)
while capturing their ``within'' correlation structure. Mishandling
this challenge can lead to an overfitted, inaccurate model. The regular
regression approach that considers each observation within an HD variable
as an independent predictor excessively increases the number of covariates
in comparison to the sample size ($p\gg n$) and ignores the correlation
between the observations. Consequently, this method may cause severe
overfitting and produce inaccurate predictions. Principle component
regression (PCR) alleviates the problem by first reducing the dimension
of both the input variables and the output. Nevertheless, PCR fails
to exploit the structure of images or point clouds\textbf{.} Furthermore,
PCR determines the principle components (PCs) of the inputs and the
outputs separately from each other without considering the interrelationship
between them. Functional data analysis, specifically the functional
regression model (FRM), has become popular in recent years due to
its built-in data reduction functionality and its ability to capture
nonlinear correlation structures \citep{liang2003relationship,Ramsay2005,yao2005functional,fan2014functional,ivanescu2015penalized,luo2017function}.
However, FRM requires a set of basis functions that is usually specified
based on domain knowledge rather than a data-driven approach. Recently,
\citet{luo2017function} proposed an approach that can combine several
profiles and scalars to predict a curve, while learning the bases
that span the input and output spaces. Nevertheless, it is not clear
how to extend this approach to other forms of data effectively. 

In the past few years, multilinear algebra (and, in particular, tensor
analysis) has shown promising results in many applications from network
analysis to process monitoring \citep{sun2006window,sapienza2015detecting,yan2015image}.
Nevertheless, only a few works in the literature use tensor analysis
for regression modeling. \citet{zhou2013tensor} has successfully
employed tensor regression using PARAFAC/CANDECOMP\textbf{ }(CP) decomposition
to estimate a scalar variable based on an image input. The CP decomposition
approximates a tensor as a sum of several rank-1 tensors \citep{kiers2000towards}.
\citet{zhou2013tensor} further extended their approach to a generalized
linear model for tensor regression in which the scalar output follows
any exponential family distribution. \citet{li2013tucker} performed
tensor regression with scalar output using Tucker decomposition. Tucker
decomposition is a form of higher order PCA that decomposes a tensor
into a core tensor multiplied by a matrix along each mode \citep{tucker1963implications}.
CP decomposition is a special case of Tucker decomposition that assumes
the same rank for all the basis matrices. \citet{yan20173DPoint}
performed the opposite regression and estimated point cloud data using
a set of scalar process variables. Recently, \citet{lock2017tensor}
developed a tensor-on-tensor regression (TOT) approach that can estimate
a tensor using a tensor input while learning the decomposition bases.
However, several limitations in TOT should be addressed. First, TOT
uses CP decomposition, which restricts both the input and output bases
to have exactly the same rank (say, $R$). This restriction may cause
over- or under-estimation when the input and the output have different
ranks. For example, when estimating an image based on few scalar inputs,
the rank of the output can be far larger than the input matrix. Second
and more importantly, this approach can only take into account a single
tensor input and cannot be used effectively when multiple sources
of input data with different dimensions and forms (e.g., a combination
of scalar, curve, and image data) are available. The extension of
TOT to multiple tensor inputs generates a significant challenge due
to the first limitation previously mentioned. Because the output and
the inputs should have the same rank, extending the TOT approach to
multiple tensor inputs as well requires all the inputs to have the
same rank (which is equal to the rank of the output). However, this
means that in certain situations, such as when scalar and image inputs
exist, one of the inputs should take the rank of the other, causing
a severe underfitting or overfitting problem. One approach to allow
multiple forms of data in TOT is to combine all the data into one
single input tensor, e.g., an image and scalar input being merged
by transforming each scalar value into a constant image. However,
this approach generates a few issues: First, it significantly increases
the size of the data, and second, it masks the separate effect of
each input on the output due to the fusion of inputs. Furthermore,
in situations in which the dimension of modes does not match (for
example, a curve input with 60 observed points and an image of size
50x50), merging data into one tensor requires a method for dimension
matching. Finally, the TOT approach fails to work on tensors of moderate
size (e.g., on the image data of size 20000 pixels used in our case
study) due to its high space complexity.

The overarching goal of this paper is to overcome the limitations
of the previous methods, such as PCR, FRMs, and TOT, by constructing
a unified regression framework that estimates a scalar, curve, image,
or structured point cloud based on a heterogeneous set of (HD) input
variables. This will be achieved by representing the output and each
group of input variables as separate tensors and by developing a multiple
tensor-on-tensor regression (MTOT). To avoid overfitting and estimating
a large number of parameters, we perform Tucker decomposition on each
group of inputs' parameters using one set of bases to expand each
of the input spaces and another set of bases to span the output space.
We obtain the input bases by performing Tucker decomposition on the
input tensors, then define a least square loss function to estimate
the decomposition coefficients and output bases. To ensure uniqueness,
we impose an orthonormality constraint over the output bases when
minimizing the loss function and show a closed-form solution for both
the bases and the decomposition coefficient in each iteration of our
algorithm. This approach not only performs dimension reduction similar
to PCR, but it also learns the output bases in accordance with the
input space. Furthermore, the use of tensors to represent the data
preserves the structure of an image or structured point cloud.

The rest of the article is organized as follows: In Section \ref{sec:Tensor-Notation},
we introduce notations and the multilinear algebra concepts used in
the paper. In Section \ref{sec:Tensor-on-Tensor}, we formulate the
multiple tensor-on-tensor regression model and illustrate the closed-form
solution for estimating the parameters. In Section \ref{sec:Performance-Evaluation-Simulation},
we describe four simulation studies. The first simulation study combines
a profile and scalar data to estimate a profile output. This simulation
study is particularly considered to compare MTOT to the available
methods in functional regression. The second and third simulation
studies contain images or point clouds either as the input or output.
The fourth simulation considers estimating a nonsmooth profile using
a set of scalar inputs. In each simulation study, we compare the performance
of the proposed method with benchmarks in terms of (standardized)
mean square prediction errors (MSPE). A case study on predicting the
overlay errors based on the wafer shape is conducted in Section \ref{sec:Case-Study}.
Finally, we summarize the paper in Section \ref{sec:Conclusion}. 

\section{Tensor Notation and Multilinear Algebra \label{sec:Tensor-Notation}}

In this section, we introduce the notations and basic tensor algebra
used in this paper. Throughout the paper, we denote a scalar by a
lower or upper case letter, e.g., $a$ or $A$; a vector by a boldface
lowercase letter and a matrix by a boldface uppercase letter, e.g.,
$\boldsymbol{a}$ and $\mathbf{A}$; and a tensor by a calligraphic
letter, e.g., $\mathcal{A}$. For example, we denote an order-$n$
tensor by $\mathcal{R\in\mathbb{R}}^{I_{1}\times I_{2}\times\cdots\times I_{n}}$,
where $I_{i}$ is the dimension of the $i^{th}$ mode of tensor $\mathcal{R}$.
We also denote a mode-$j$ matricization of tensor $\mathcal{R}$
as $\boldsymbol{R}_{\left(j\right)}\in\mathcal{\mathbb{R}}^{I_{j}\times I_{-j}}$,
whose columns are the mode-$j$ fibers of the corresponding tensor
$\mathcal{R}$, and $I_{-j}=I_{1}\times I_{2}\times\cdots\times I_{j-1}\times I_{j+1}\times\cdots\times I_{n}$.
We also define a more general matricization of a tensor $\mathcal{T}\in\mathbb{R}^{P_{1}\times\cdots\times P_{l}\times Q_{1}\times\cdots\times Q_{d}}$
as follows: Let $\mathbb{I}=\left\{ I_{1},I_{2},\cdots,I_{l}\right\} $
and $\mathbb{Q}=\left\{ Q_{1},Q_{2},\cdots,Q_{d}\right\} $ be two
sets that partition the set $\left\{ I_{1},I_{2},\cdots,I_{l},Q_{1},Q_{2},\cdots,Q_{d}\right\} $,
which contains the dimensions of the modes of the tensor $\mathcal{\mathcal{T}}$.
Then, the matricized tensor is specified by $\boldsymbol{T}_{\left(\mathbb{I}\times\mathbb{Q}\right)}\in\mathbb{R}^{P\times Q}$,
where $P=\prod_{i=1}^{l}P_{i}$ and $Q=\prod_{i=1}^{d}Q_{i}$, and
\[
\left(\boldsymbol{T}_{\left(\mathbb{I}\times\mathbb{Q}\right)}\right)_{ij}=\mathcal{T}_{p_{1}\cdots p_{l}q_{1}\cdots q_{d}},
\]
where $i=1+\sum_{r=1}^{l}\prod_{n=1}^{r}P_{n}\left(p_{n}-1\right)$
and $j=1+\sum_{r=1}^{d}\prod_{n=1}^{r}Q_{n}\left(q_{n}-1\right)$.
For simplicity of notation, we will denote $\boldsymbol{T}_{\left(\mathbb{I}\times\mathbb{Q}\right)}$
as \textbf{$\boldsymbol{T}$}.

The Frobenius norm of a tensor $\mathcal{R}$ can be defined as the
Frobenius norm of any matricized tensor, e.g., $||\mathcal{R}||_{F}^{2}=||\boldsymbol{R}_{(1)}||_{F}^{2}$.
The mode-$j$ product of a tensor $\mathcal{R}$ by a matrix $\boldsymbol{A}\in\mathbb{R}^{L\times I_{j}}$
is a tensor in $\mathcal{\mathbb{R}}^{I_{1}\times I_{2}\times\cdots\times I_{j-1}\times L\times I_{j+1}\times\cdots\times I_{n}}$
and is defined as 
\[
\left(\mathcal{R}\times_{j}\boldsymbol{A}\right)_{i_{1},i_{2},\cdots,i_{j-1},l,i_{j+1},\cdots,i_{K}}=\sum_{i_{j}=1}^{I_{j}}\mathcal{R}_{i_{1},\cdots,i_{j},\cdots,i_{K}}\boldsymbol{A}_{l,i_{j}}.
\]
The Tucker decomposition of a tensor $\mathcal{R}$ decomposes the
tensor into a core tensor $\mathcal{C}\in\mathbb{R}^{P_{1}\times P_{2}\times\cdots\times P_{n}}$
and $n$ orthogonal matrices $\boldsymbol{U}_{i}\in\mathbb{R}^{I_{i}\times P_{i}}\,\left(i=1,2,\cdots,n\right)$
so that $\mathcal{R}=\mathcal{C}\times_{1}\boldsymbol{U}_{1}\times_{2}\boldsymbol{U}_{2}\times_{3}\cdots\times_{n}\boldsymbol{U}_{n}$.
The dimensions of the core tensor $\mathcal{C}$ is smaller than $\mathcal{A}$,
i.e., $P_{j}\leq I_{j}\,\left(j=1,2,\cdots,n\right)$. Furthermore,
the Kronecker product of two matrices $\boldsymbol{A}\in\mathbb{R}^{m\times n}$
and $\boldsymbol{B}\in\mathbb{R}^{r\times s}$ is denoted as $\boldsymbol{A}\otimes\boldsymbol{B}\in\mathbb{R}^{mr\times ns}$
and is obtained by multiplying each element of matrix $\boldsymbol{A}$
to the entire matrix $\boldsymbol{B}$:
\[
\boldsymbol{A}\otimes\boldsymbol{B}=\left[\begin{array}{ccc}
a_{11}\boldsymbol{B} & ... & a_{1n}\boldsymbol{B}\\
\vdots & \ddots & \vdots\\
a_{m1}\boldsymbol{B} & ... & a_{mn}\boldsymbol{B}
\end{array}\right].
\]
We link the tensor multiplication with the Kronecker product using
the Proposition \ref{prop:proposition1}.
\begin{prop}
\label{prop:proposition1}Let $\boldsymbol{U}_{i}\in\mathbb{R}^{P_{i}\times\tilde{P_{i}}}\left(i=1,\cdots,l\right)$
and $\boldsymbol{V}_{i}\in\mathbb{R}^{Q_{i}\times\tilde{Qi}}\,\left(i=1,\cdots,d\right)$,
and let $\mathcal{T}\in\mathbb{R}^{P_{1}\times\cdots\times P_{l}\times Q_{1}\times\cdots\times Q_{d}}$
and $\mathcal{C}\mathbb{\in R}^{\tilde{P_{1}}\times\cdots\times\tilde{P_{l}}\times\tilde{Q_{1}}\times\cdots\times\tilde{Q}_{d}}$,
then
\begin{align*}
 & \mathcal{T}=\mathcal{C}\times_{1}\boldsymbol{U}_{1}\times\boldsymbol{U}_{2}\times_{3}\cdots\times_{l}\boldsymbol{U}_{l}\times_{l+1}\boldsymbol{V}_{1}\times_{l+2}\cdots\times_{l+d}\boldsymbol{V}_{d}\iff\\
 & \boldsymbol{T}=\left(\boldsymbol{U}_{l}\otimes\boldsymbol{U}_{l-1}\otimes\cdots\otimes\boldsymbol{U}_{1}\right)\boldsymbol{C}\left(\boldsymbol{V}_{d}\otimes\boldsymbol{V}_{d-1}\otimes\cdots\otimes\boldsymbol{V}_{1}\right)^{\mathtt{T}},
\end{align*}
where $\boldsymbol{C}\in\mathbb{R}^{\tilde{P}\times\tilde{Q}}$ is
an unfold of the core tensor $\mathcal{C}$ with $\tilde{P}=\prod_{j=1}^{l}\tilde{P_{j}}$
and $\tilde{Q}=\prod_{j=1}^{d}\tilde{Q}_{j}$. 
\end{prop}

The proof of this proposition can be found in \citep{kolda2006multilinear}.
Finally, the contraction product of two tensors $\mathcal{B}\in\mathbb{R}^{P_{1}\times\cdots\times P_{l}\times Q_{1}\times\cdots\times Q_{d}}$
and $\mathcal{X}\in\mathbb{R}^{P_{1}\times\cdots\times P_{l}}$ is
denoted as $\mathcal{X}*\mathcal{B}\in\mathbb{R}^{Q_{1}\times\cdots\times Q_{d}}$
and is defined as 
\[
\left(\mathcal{X}*\mathcal{B}\right)_{q_{1}\cdots q_{d}}=\sum_{p_{1},\cdots,p_{l}}\mathcal{X}_{p_{1},\cdots,p_{l}}\mathcal{B}_{p_{1},\cdots,p_{l},q_{1},\cdots,q_{d}}.
\]

\section{Multiple Tensor-on-Tensor Regression Framework \label{sec:Tensor-on-Tensor}}

In this section, we introduce the multiple tensor-on-tensor (MTOT)
framework as an approach for integrating multiple sources of data
with different dimensions and forms to model a process. Assume a set
of training data of size $M$ is available and includes response tensors
$\mathcal{Y}_{i}\in\mathbb{R}^{Q_{1}\times\cdots\times Q_{d}}\,\left(i=1,\cdots,M\right)$
and input tensors $\mathcal{X}_{ji}\in\mathbb{R}^{P_{j1}\times\cdots\times P_{jl_{j}}}\,\left(i=1,\cdots,M;\,j=1,\cdots,p\right)$,
where $p$ is the number of inputs. The goal of MTOT is to model the
relationship between the input tensors and the response using the
linear form
\begin{equation}
\mathcal{Y}_{i}=\sum_{j=1}^{p}\mathcal{X}_{ji}*\mathcal{B}_{j}+\mathcal{E}_{i},\,i=1,\cdots M,\label{eq:simple regression-1}
\end{equation}
where $\mathcal{B}_{j}\in\mathbb{R}^{P_{j1}\times\cdots\times P_{jl_{j}}\times Q_{1}\times\cdots\times Q_{d}}$
is the model parameter to be estimated and $\mathcal{E}_{i}$ is an
error tensor whose elements are from a random process. To achieve
a more compact representation of the model (\ref{eq:simple regression-1}),
we can combine tensors $\mathcal{Y}_{i}$, $\mathcal{X}_{ji}$, and
$\mathcal{E}_{i}$ $\left(i=1,\cdots,M\right)$ into one-mode larger
tensors $\mathcal{Y}\in\mathbb{R}^{M\times Q_{1}\times\cdots\times Q_{d}}$,
$\mathcal{X}_{j}\in\mathbb{R}^{M\times P_{j1}\times P_{j2}\times\cdots\times P_{jl_{j}}}\,\left(j=1,2,\cdots,p\right)$,
and $\mathcal{E}\in\mathbb{R}^{M\times Q_{1}\times\cdots\times Q_{d}}$
and write 
\begin{equation}
\mathcal{Y}=\sum_{j=1}^{p}\mathcal{X}_{j}*\mathcal{B}_{j}+\mathcal{E}.\label{eq:multipleTensorRegressionFormula}
\end{equation}
The matricization of (\ref{eq:multipleTensorRegressionFormula}) gives
\begin{equation}
\boldsymbol{Y}_{\left(1\right)}=\sum_{j=1}^{p}\boldsymbol{X}_{j\left(1\right)}\boldsymbol{B}_{j}+\boldsymbol{E}_{\left(1\right)},\label{eq:multiple regression matricization}
\end{equation}
where $\boldsymbol{Y}_{\left(1\right)}$ and $\boldsymbol{X}_{j\left(1\right)}$
are mode-1 unfolding of tensors $\mathcal{Y}$ and $\mathcal{X}_{j}$,
respectively, and the first mode corresponds to the sample mode. $\boldsymbol{B}_{j}\in\mathbb{R}^{P_{j}\times Q}$
is an unfold of tensor $\mathcal{B}_{j}$ with $P_{j}=\prod_{k=1}^{l_{j}}P_{jk}$
and $Q=\prod_{k=1}^{d}Q_{k}$. It is intuitive that the parameters
of (\ref{eq:multiple regression matricization}) can be estimated
by minimizing the mean squared loss function $L=||\boldsymbol{Y}_{\left(1\right)}-\sum_{j=1}^{p}\boldsymbol{X}_{j\left(1\right)}\boldsymbol{B}_{j}||_{F}^{2}$.
However, this requires estimating $\sum_{j=1}^{p}\prod_{i=1}^{l_{j}}P_{ji}\prod_{k=1}^{d}Q_{k}$
parameters. For example, in the situation in which\textbf{ }$p=1$,
minimizing the loss function gives a closed-form solution $\hat{\boldsymbol{B}}=\left(\boldsymbol{X}_{\left(1\right)}^{\mathtt{T}}\boldsymbol{X}_{\left(1\right)}\right)^{-1}\boldsymbol{X}_{\left(1\right)}^{\mathtt{T}}\boldsymbol{Y}_{\left(1\right)}$
that requires estimating $\prod_{i=1}^{l}P_{i}\prod_{j=1}^{d}Q_{j}$
parameters. Estimating such a large number of parameters is prone
to severe overfitting and is often intractable. In reality, due to
the structured correlation between $\mathcal{X}_{j}$ and $\mathcal{Y}$,
we can assume that the parameter $\mathcal{B}_{j}$ lies in a much
lower dimensional space and can be expanded using a set of basis matrices
via a tensor product. That is, for each $\mathcal{B}_{j}\,\left(j=1,\cdots,p\right)$,
we can write
\begin{equation}
\mathcal{B}_{j}=\mathcal{C}_{j}\times_{1}\boldsymbol{U}_{j1}\times_{2}\boldsymbol{U}_{j2}\times_{3}\cdots\times_{l_{j}}\boldsymbol{U}_{jl_{j}}\times_{l_{j}+1}\boldsymbol{V}_{1}\times_{l_{j}+2}\cdots\times_{l_{j}+d}\boldsymbol{V}_{d},\label{eq:MR-parameter expansion}
\end{equation}
where $\mathcal{C}_{j}\mathbb{\in R}^{\tilde{P_{j1}}\times\cdots\times\tilde{P_{jl_{j}}}\times\tilde{Qi}\times\cdots\times\tilde{Q}_{d}}$
is a core tensor with $\tilde{P_{ji}}\ll P_{ji}\,\left(j=1,\cdots,p;\,i=1,\cdots l_{j}\right)$
and $\tilde{Q}_{i}\ll Q_{i}\,\left(i=1,\cdots,d\right)$; $\left\{ \boldsymbol{U}_{ji}:j=1,\cdots p;\,i=1,\cdots,l_{j}\right\} $
is a set of bases that spans the $j^{th}$ input space; and $\left\{ \boldsymbol{V}_{i}:\,i=1,\cdots,d\right\} $
is a set of bases that spans the output space. With this low-dimensional
representation, the estimation of $\mathcal{B}_{j}$ reduces to learning
the core tensor $\mathcal{C}_{j}$ and the basis matrices $\boldsymbol{U}_{ji}$
and $\boldsymbol{V}_{i}$. In this paper, we allow $\boldsymbol{U}_{ji}$
to be learned directly from the input spaces. Two important choices
of $U_{ji}$ are truncated identity matrices (i.e., no transformation
on the inputs) or the bases obtained from Tucker decomposition of
the input tensor $\mathcal{X}_{j}$, i.e., 
\[
\left\{ \mathcal{D}_{j},\boldsymbol{U}_{j1},\cdots,\boldsymbol{U}_{jl_{j}}\right\} =\arg\min_{\mathcal{D}_{j},\left\{ \boldsymbol{U}_{ji}\right\} }\left|\left|\mathcal{X}_{j}-\mathcal{D}_{j}\times_{1}\boldsymbol{U}_{j1}\times\cdots\times_{l_{j}}\boldsymbol{U}_{jl}\right|\right|_{F}^{2}.
\]
In a special case that an input tensor is a matrix, the bases are
the principle components (PCs) of that input if one uses Tucker decomposition.
Allowing $\boldsymbol{U}_{ji}\,\left(j=1,\cdots,p;\,i=1,\cdots l_{j}\right)$
to be selected is reasonable because $\mathcal{X}_{j}$ is an independent
variable, and its basis matrices can be obtained separately from the
output space. Furthermore, learning the core tensors $\mathcal{C}_{j}\,\left(j=1,\cdots,p\right)$
and the bases $\boldsymbol{V}_{i}\,\left(i=1,\cdots,d\right)$ provides
a sufficient degree of freedom to learn $\mathcal{B}_{j}$. Next,
we iteratively estimate the core tensors $\mathcal{C}_{j}$ and the
basis matrices $\boldsymbol{V}_{i}$ by solving the following optimization
problem: 
\begin{align}
 & \left\{ \mathcal{C}_{j},\boldsymbol{V}_{1},\cdots,\boldsymbol{V}_{d}\right\} =\arg\min\left|\left|\boldsymbol{Y}_{\left(1\right)}-\sum_{j=1}^{p}\boldsymbol{X}_{j\left(1\right)}\boldsymbol{B}_{j}\right|\right|_{F}^{2},\,s.t.\nonumber \\
 & \mathcal{B}_{j}=\mathcal{C}_{j}\times_{1}\boldsymbol{U}_{j1}\times_{2}\boldsymbol{U}_{j2}\times_{3}\cdots\times_{l_{j}}\boldsymbol{U}_{jl_{j}}\times_{l_{j}+1}\boldsymbol{V}_{1}\times_{l_{j}+2}\cdots\times_{l_{j}+d}\boldsymbol{V}_{d},\nonumber \\
 & \boldsymbol{V}_{i}^{\mathtt{T}}\boldsymbol{V}_{i}=I_{\tilde{Qi}}\,\left(i=1,\cdots,d\right),\label{eq:multipleTOT-objectiveFunction}
\end{align}
where $\boldsymbol{I}_{\tilde{Qi}}$ is a $\tilde{Qi}\times\tilde{Qi}$
identity matrix. The first constraint ensures that the tensor of parameters
is low-rank, and the orthogonality constraint $\boldsymbol{V}_{i}^{\mathtt{T}}\boldsymbol{V}_{i}=\boldsymbol{I}_{\tilde{Qi}}$
ensures the uniqueness of both the bases and the core tensors when
the problem is identifiable. In general, the problem of estimating
functional data through a set of functions may not be identifiable
under some conditions. That is, assuming $p=1$, one can find $\boldsymbol{B}\neq\overline{\boldsymbol{B}}$,
such that $\boldsymbol{B}X=\overline{\boldsymbol{B}}X$, i.e., $\boldsymbol{B}$
and $\overline{\boldsymbol{B}}$ both estimate same mean value for
the output. \citet{he2000extending,chiou2004functional,lock2017tensor}
discuss the identifiability problem in functional and tensor regression.
Because the main purpose of this paper is to estimate and predict
the output, we do not discuss the identifiability issue here, as learning
any correct set of parameters $\left\{ \boldsymbol{B}_{k}:\,k=1,\cdots,p\right\} $
will eventually lead to the same estimation of the output. 

In order to solve (\ref{eq:multipleTOT-objectiveFunction}), we combine
the alternating least square (ALS) approach with the block coordinate
decent (BCD) method (designated by ALS-BCD). The advantages of ALS
algorithms that lead to their widespread use are conceptual simplicity,
noise robustness, and computational efficiency \citep{sharan2017orthogonalized}.
In tensor decomposition and regression, due to the non-convex nature
of the problem, finding the global optimum is often intractable, and
it is well-known that the ALS algorithm also has no global convergence
guarantee and may be trapped in a local optima \citep{kolda2006multilinear,sharan2017orthogonalized}.
However, ALS has shown great promise in the literature for solving
tensor decomposition and regression applications with satisfying results.
To be able to employ ALS-BCD, we first demonstrate Proposition \ref{prop:MR-core tensor estimate}:
\begin{prop}
\label{prop:MR-core tensor estimate}When $\boldsymbol{U}_{ki}\,\left(k=1,\cdots,p;\,i=1,2,\cdots,l_{j}\right)$
,$\boldsymbol{V}_{i}\,\left(i=1,2,\cdots,d\right)$, and $\text{\ensuremath{\mathcal{C}}}_{k}\,\left(k\neq j\right)$
are known, a reshaped form of the core tensor $\text{\ensuremath{\mathcal{C}}}_{j}$
can be estimated as 
\begin{equation}
\text{\ensuremath{\mathcal{\tilde{C}}}}_{j}=\mathcal{R}_{j}\times_{1}\left(\boldsymbol{Z}_{j}^{\mathtt{T}}\boldsymbol{Z}_{j}\right)^{-1}\boldsymbol{Z}_{j}^{\mathtt{T}}\times_{2}\left(\boldsymbol{V}_{1}^{\mathtt{T}}\boldsymbol{V}_{1}\right)^{-1}\boldsymbol{V}_{1}^{\mathtt{T}}\times_{3}\left(\boldsymbol{V}_{2}^{\mathtt{T}}\boldsymbol{V}_{2}\right)^{-1}\boldsymbol{V}_{2}^{\mathtt{T}}\cdots\times_{d+1}\left(\boldsymbol{V}_{d}^{\mathtt{T}}\boldsymbol{V}_{d}\right)^{-1}\boldsymbol{V}_{d}^{\mathtt{T}},\label{eq:SolutionToRegression-1}
\end{equation}
where $\boldsymbol{Z}_{j}=\boldsymbol{X}_{j\left(1\right)}(\boldsymbol{U}_{jl}\otimes\boldsymbol{U}_{jl-1}\otimes\cdots\otimes\boldsymbol{U}_{j1})$
and $\mathcal{R}_{j}=\mathcal{Y}-\sum_{i\neq j}^{p}\mathcal{B}_{j}*\mathcal{X}_{j}$.
Note that $\text{\ensuremath{\mathcal{\tilde{C}}}}_{j}$ has fewer
modes ($d+1$) than the original core tensor $\mathcal{C}_{j}$ in
(\ref{eq:MR-parameter expansion}), but it can be transformed into
$\mathcal{C}$ by a simple reshape operation.
\end{prop}

The simplified proof of this proposition is given in Appendix A. Furthermore,
if $\boldsymbol{V}_{i}$s are orthogonal, the core tensor can be obtained
efficiently by the tensor product as
\[
\text{\ensuremath{\mathcal{\tilde{C}}}}_{j}=\mathcal{R}_{j}\times_{1}\left(\boldsymbol{Z}_{j}^{\mathtt{T}}\boldsymbol{Z}_{j}\right)^{-1}\boldsymbol{Z}_{j}^{\mathtt{T}}\times_{2}\boldsymbol{V}_{1}^{\mathtt{T}}\times_{3}\boldsymbol{V}_{2}^{\mathtt{T}}\cdots\times_{d+1}\boldsymbol{V}_{d}^{\mathtt{T}}.
\]
Note that in the situations in which sparsity of the core tensor is
of interest, one can include a lasso penalty over the core tensor,
and use numerical algorithms (e.g., Iterative Shrinkage-Thresholding
Algorithm \citep{beck2009fast}) to solve the problem. Furthermore,
one can estimate the basis matrices $V_{i}$ as follows:
\begin{prop}
\label{prop:MR-basis estimate}With known $\mathcal{C}_{j}$, $\boldsymbol{U}_{ji}$,
and $\boldsymbol{V}_{k}\,\left(k\neq i\right)$, we can solve $\boldsymbol{V}_{i}$
by
\[
\boldsymbol{V}_{i}=\boldsymbol{R}\boldsymbol{W}^{\mathtt{T}},
\]
where $\boldsymbol{R}$ and $\boldsymbol{W}$ are obtained from the
singular value decomposition of $\boldsymbol{Y}_{\left(i\right)}\boldsymbol{S}^{\mathtt{T}}$,
where $\boldsymbol{S}=\sum_{j=1}^{p}\boldsymbol{S}_{j}$ and $\boldsymbol{S}_{j}=\tilde{\boldsymbol{C}}_{j\left(i\right)}\left(\boldsymbol{V}_{d}\otimes\cdots\otimes\boldsymbol{V}_{i+1}\otimes\boldsymbol{V}_{i-1}\cdots\otimes\boldsymbol{V}_{1}\otimes\boldsymbol{Z}_{j}\right)^{T}$;
and $\tilde{\boldsymbol{C}}_{j\left(i\right)}$ is the mode-i matricization
of tensor $\text{\ensuremath{\mathcal{\tilde{C}}}}_{j}$. Note that
$\boldsymbol{R}$ is truncated.
\end{prop}

The simplified proof of this proposition is shown in Appendix B. First
note that we do not require calculation of the Kronecker product $\boldsymbol{V}_{d}\otimes\cdots\boldsymbol{V}_{i+1}\otimes\boldsymbol{V}_{i-1}\otimes\boldsymbol{V}_{1}\otimes\boldsymbol{Z}$
explicitly to find $\boldsymbol{Y}_{\left(i\right)}\boldsymbol{S}^{\mathtt{T}}$.
In real implementation, we can use Proposition \ref{prop:proposition1}
to calculate the complete matrix using tensor products efficiently.
Second, unlike the principle component regression (PCR) in which the
principle components of the output are learned independent of the
inputs, the estimated basis matrices $\boldsymbol{V}_{i}\,\left(i=1,\cdots,d\right)$
directly depend on the input tensors, ensuring correlation between
the bases and inputs. By combining Propositions \ref{prop:MR-core tensor estimate}
and \ref{prop:MR-basis estimate}, Algorithm \ref{alg:MR-Estimation-procedure algorithm}
summarizes the estimation procedure for multiple tensor-on-tensor
regression. This algorithm, in fact, combines the block coordinate
decent (BCD) algorithm with the ALS algorithm. 
\begin{algorithm}
\begin{algorithmic}[1]
\algdef{SE}[DOWHILE]{Do}{doWhile}{\algorithmicdo}[1]{\algorithmicwhile\ #1}%
\State{Initilize $\mathcal{C}_j$ for all $j$}
\State{Estimate $U_{ji}$ using Tucker decomposition of $\mathcal{X}_j$ for all $i$ and $j$}
\State{Initilize $V_i$ for all $i$}
\State{Compute $B_{j}$ for all $j$ and set $w_{0} =\left|\left|Y_{\left(1\right)}-\sum_{j=1}^{p}X_{j\left(1\right)}B_{j}\right|\right|_{F}^{2}$}
\Do
\State{Estimate $\mathcal{C}_j$ for all $j = 1: p$ using Proposition \ref{prop:MR-core tensor estimate}}
\State{Estimate $V_i$ for all $i = 1: d$ using Proposition \ref{prop:MR-basis estimate}}
\State{Compute $B_{j}$ for all $j$ and set $w_{k} =\left|\left|Y_{\left(1\right)}-\sum_{j=1}^{p}X_{j\left(1\right)}B_{j}\right|\right|_{F}^{2}$}
\doWhile{$|w_{k+1}-w_{k}| > \epsilon $}
\end{algorithmic}

\caption{Estimation procedure for multiple tensor-on-tensor regression \label{alg:MR-Estimation-procedure algorithm}}
\end{algorithm}

\subsection{Selection of tuning parameters}

The proposed approach requires the selection of the values $\tilde{P}_{ji}$~$\left(j=1,2,\cdots,p;i=1,2,\cdots,l_{j}\right)$
and $\tilde{Q}_{k}\,\left(k=1,\cdots,d\right)$. For simplicity and
practicality, we assume that for each predictor $\mathcal{X}_{j}$
and the response $\mathcal{Y}$, the rank is fixed, i.e., $\tilde{P}_{j1}=\tilde{P}_{j2}=\cdots=\tilde{P}_{jl_{j}}=\tilde{P}_{j}$
and $\tilde{Q}_{k}=\tilde{Q}$. As a result, we only need to select
$p+1$ parameters. For this purpose, we use the k-fold cross-validation
method on a $\left(p+1\right)$-D grid of parameters $\left(\tilde{P}_{1},\tilde{P}_{2},\cdots,\tilde{P}_{p},\tilde{Q}\right)$
and find the tuple of parameters that minimizes the mean squared error.
As a result, we should define a grid over the rank values. This is
achieved as following: First, we unfold each tensor $\mathcal{X}_{j}\,\left(j=1,2,\cdots,p\right)$
and $\mathcal{Y}$ along their first mode. Next, we find the rank
of each unfolded matrix, denoted as $R_{x_{1}},R_{x_{2}},\cdots,R_{x_{p}}$
and $R_{y}$. Then, for each $\mathcal{X}_{j}\,\left(j=1,2,\cdots,p\right)$
and $\mathcal{Y}$, we select the tuning parameters from $\mathcal{P}_{j}=\left\{ 1,\left\lceil \frac{R_{x_{j}}}{2^{log_{2}R_{x_{j}}-1}}\right\rceil ,\cdots\left\lceil \frac{R_{x_{j}}}{2^{2}}\right\rceil ,\left\lceil \frac{R_{x_{j}}}{2}\right\rceil ,R_{x_{j}}\right\} $
and $\mathcal{Q}=\left\{ 1,\left\lceil \frac{R_{y}}{2^{log_{2}R_{y}-1}}\right\rceil ,\cdots\left\lceil \frac{R_{y}}{2^{2}}\right\rceil ,\left\lceil \frac{R_{y}}{2}\right\rceil ,R_{y}\right\} $.
Next, for each tuple $\left(\tilde{P}_{1},\tilde{P}_{2},\cdots,\tilde{P}_{p},\tilde{Q}\right)\in\mathcal{P}_{1}\times\mathcal{P}_{2}\times\cdots\times\mathcal{P}_{p}\times\mathcal{Q}$,
we calculate the average sum square error (RSS) and take the one that
minimizes the RSS. For all studies in the next sections, we perform
five-fold CV. 

\section{Performance Evaluation Using Simulation \label{sec:Performance-Evaluation-Simulation}}

This section contains two parts. In the first part, we only consider
curve-on-curve regression and compare our proposed method to the function-on-function
regression approach proposed by \citet{luo2017function}, designated
as \textit{sigComp}. The reason we compare our approach to sigComp
is that sigComp can handle multiple functional inputs (curves) and
learn the basis functions similar to our approach. In the second part,
we conduct a set of simulation studies to evaluate the performance
of the proposed method when the inputs or outputs are in the form
of images, structured point clouds, or curves with jumps. In this
part, we compare the proposed method with two benchmarks: 1) The first
benchmark is the TOT approach proposed by \citet{lock2017tensor},
which can roughly be viewed as a general form of sigComp. Because
this approach can only handle a single input tensor, when multiple
inputs exist we perform a transformation to merge the inputs into
one single tensor. 2) The second benchmark is based on principle component
regression (PCR) similar to a benchmark considered in \citep{fan2014functional}.
In this approach, we first matricize all the input and output tensors,
then perform principle component analysis to reduce the dimension
of the problem by computing the PCA scores of the first few principle
components that explain at least $v$ percent of the variation in
the data. Next, we perform linear regression between the low-dimensional
PC score of both inputs and output. More formally, let $\boldsymbol{X}_{j\left(1\right)}\in\mathbb{R}^{M\times P_{j}}$
and $\boldsymbol{Y}_{\left(1\right)}\in\mathbb{R}^{M\times Q}$ denote
the mode-1 matricization of the inputs and output, and $\boldsymbol{X}=\left[\boldsymbol{X}_{1\left(1\right)},\boldsymbol{X}_{2\left(1\right)},\cdots,\boldsymbol{X}_{p\left(1\right)}\right]$
be a concatenation of all the input matrices. We first compute the
first $G_{x}$ and $G_{y}$ principle components of $X$ and the response
$\boldsymbol{Y}_{\left(1\right)}$. Next, the PC scores of the input
$\boldsymbol{X}$ are calculated (a matrix in $\mathbb{R}^{M\times G_{x}}$)
and are used to predict the matrix of the scores of the response function
(a matrix in $\mathbb{R}^{M\times G_{y}}$). Then, given the PC scores
of the new inputs, we use the fitted regression model to predict the
response scores. Finally, we multiply the predicted response scores
by the $G_{y}$ principle components to obtain the original responses.
The number of principle components $G_{x}$ and $G_{y}$ can be identified
through a cross-validation procedure. In this paper, instead of cross-validating
over $G_{x}$ and $G_{y}$ directly, we perform CV over the percentage
of variation the PCs explain, i.e., $v$. For this purpose, we take
the value of $v$ from $\{85\%,\,90\%,\,95\%,\,99\%,\,99.5\%\}$ and
take the $v$ that minimizes the CV error. The standardized mean square
prediction error (SMSPE) is used as a performance measure to compare
the proposed method with the benchmarks. The SMSPE is defined as $SMSPE=\frac{||\mathcal{Y}-\hat{\mathcal{Y}}||_{F}^{2}}{||\mathcal{Y}||_{F}^{2}}.$ 

\subsection{Simulation studies for curve-on-curve regression}

\begin{figure}
\begin{centering}
\subfloat[]{\includegraphics[width=0.35\columnwidth]{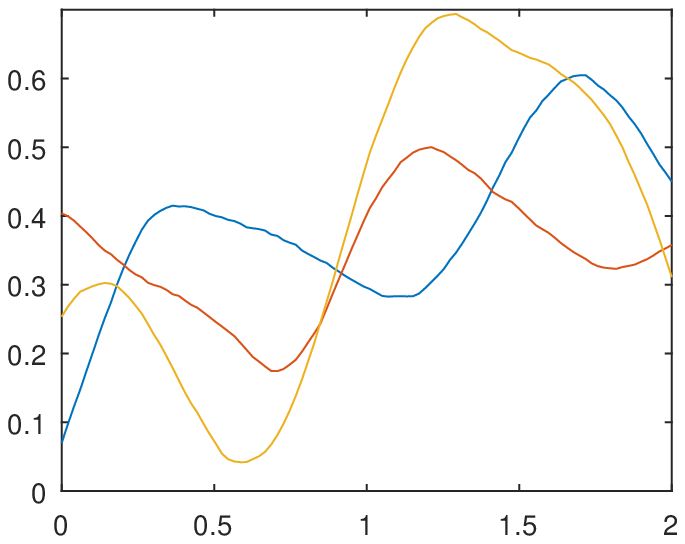}

}\subfloat[]{\includegraphics[width=0.35\columnwidth]{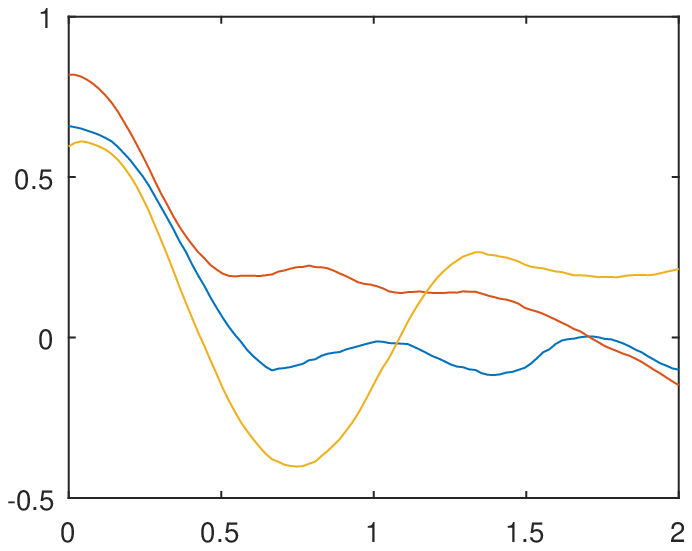}}
\par\end{centering}
\caption{Example of the predictors when (a) $p=3$, $\rho_{c}=0$ and (b) $p=3$,$\rho_{c}=0.5$.\label{fig:Example-of-the_predictors_sim0}}
\end{figure}

In this simulation, we consider multiple functional (curve) predictors
and multiple scalar predictors similar to the simulation study in
\citep{luo2017function}. We first randomly generate $\left(\boldsymbol{B}_{1},\boldsymbol{B}_{2},\cdots,\boldsymbol{B}_{p}\right)$
as follows: 
\[
\boldsymbol{B}_{i}\left(s,t\right)=\frac{1}{p^{2}}\left[\gamma_{1i}\left(t\right)\psi_{1i}\left(s\right)+\gamma_{2i}\left(t\right)\psi_{2i}\left(s\right)+\gamma_{3i}\left(t\right)\psi_{3i}\left(s\right)\right],
\]
where $\gamma_{ki}\left(t\right)$ and $\psi_{ki}\left(s\right)\,\left(k=1,\,2,\,3;\,i=1,\cdots,p\right)$
are Gaussian processes with covariance function $\Sigma_{1}\left(z,z'\right)=\left(1+20\left|z-z'\right|+\frac{1}{3}\left(20\left|z-z'\right|\right)^{2}\right)e^{-20\left|z-z'\right|}$.
Next, we generate $p=1,\,3,\,6,$ functional predictors using the
following procedure: Let $S$ be a $p\times p$ matrix with the $\left(i,j\right)^{th}$
element equal to $\rho_{c}=0,\,0.5$ for $i\neq j$ and equal to one
for diagonal elements. Next, we decompose $\boldsymbol{S}=\boldsymbol{\Delta}\boldsymbol{\Delta}^{\mathtt{T}}$,
where $\boldsymbol{\Delta}$ is a $p\times p$ matrix and generate
a set of curves $w_{1},\,w_{2},\cdots,\,w_{p}$ using a Gaussian process
with covariance function $\Sigma_{2}\left(z,z'\right)=e^{-\left(2\left|z-z'\right|\right)^{2}}$.
Finally, we generate the predictors at any given point $s$ as 
\[
\left(x_{1}\left(s\right),\cdots,x_{p}\left(s\right)\right)=\left(w_{1}\left(s\right),w_{2}\left(s\right),\cdots,w_{p}\left(s\right)\right)\boldsymbol{\Delta}^{\mathtt{T}}.
\]
With this formulation, each curve of $x_{1}\left(s\right),\cdots,x_{p}\left(s\right)$
is a Gaussian process with covariance function $\Sigma_{2}\left(s,s'\right)$,
and for each $s$, the vector $\left(x_{1}\left(s\right),\cdots,x_{p}\left(s\right)\right)$
is a multivariate normal distribution with covariance \textbf{$\boldsymbol{S}$}.
When $\rho_{c}=0$, this vector becomes an independent vector of normally
distributed variables. Figure \ref{fig:Example-of-the_predictors_sim0}
illustrates examples of the predictors when $p=3$ for $\rho_{c}=0,\,0.5$.
We also generate the scalar predictors $\left(u_{1},\cdots,u_{5}\right)$
from a multivariate normal distribution with mean vector zero and
the covariance matrix with diagonal elements equal to $1$ and off-diagonal
elements equal to $0.5$. The coefficients of the scalar variables
denoted by $\alpha_{i}\left(t\right)\,\left(i=1,\cdots,5\right)$
are generated from a Gaussian process with covariance function $\Sigma\left(t,t'\right)=e^{\left\{ -5\left|t-t'\right|\right\} ^{2}}.$
Finally, we generate the response curves as 
\[
y\left(t\right)=\sum_{i=1}^{5}\alpha_{i}\left(t\right)u_{i}+\sum_{i=1}^{p}\intop\boldsymbol{B}_{i}\left(s,t\right)x_{i}\left(s\right)ds+\epsilon\left(t\right),
\]
where $\epsilon\left(t\right)$ is generated from a normal distribution
with zero mean and $\sigma^{2}=0.1$. We generate all of the input
and output curves over $0<s<2$ and $0<t<1$ and take the samples
over an equidistant grid of size $100$.

\begin{figure}
\begin{centering}
\includegraphics[bb=20bp 0bp 320bp 255bp,clip,width=0.6\columnwidth]{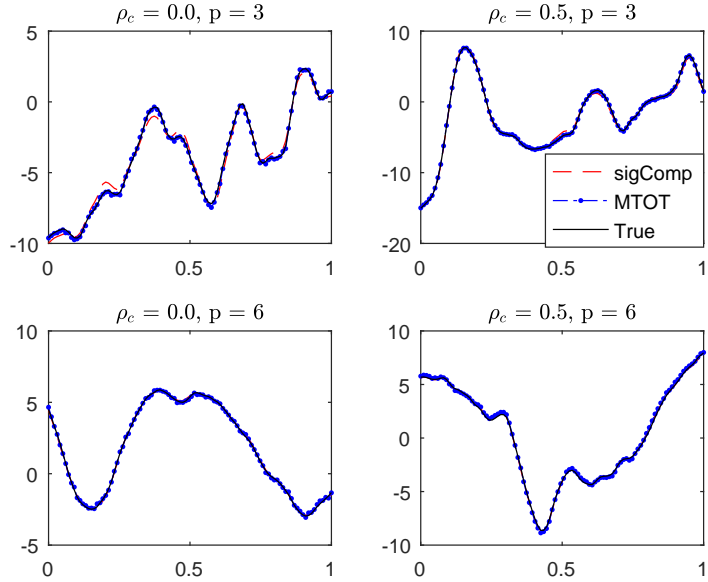}
\par\end{centering}
\caption{Prediction examples of sigComp and MTOT for different $p$ and $\rho_{c}$.\label{fig:Prediction-examples_GPsim}}

\end{figure}

For each combination of $\left(p,\rho_{c}\right)$, we compare the
performance of the proposed method with the methods in \citep{luo2017function}
based on the mean square prediction error (MSPE) and the mean square
estimation error (MSEE). We do not compare our approach to PCR in
this simulation because sigComp has already demonstrated superiority
over PCR in simulation studies in \citep{luo2017function}. We implement
the sigComp benchmark method using the R package \textit{FRegSigCom}
in which we use 50 spline bases for both the inputs and output and
default convergence tolerance. To calculate the MSPE and MSEE, we
first generate a sample data of size $M_{train}=400$ that is used
to learn the model parameters. Next, we generate a testing sample
of size $M_{test}=100$ and calculate MSPE as
\[
MSPE=\frac{1}{M_{test}}\sum_{j=1}^{M_{test}}\left(\frac{1}{100}\sum_{i=1}^{100}\left(y_{j}^{test}\left(t_{i}\right)-\hat{y}_{j}^{test}\left(t_{i}\right)\right)^{2}\right)
\]
and 
\[
MSEE=\frac{1}{M_{test}}\sum_{j=1}^{M_{test}}\frac{1}{100}\sum_{i=1}^{100}\left(y_{j}^{test}\left(t_{i}\right)-\epsilon\left(t_{i}\right)-\hat{y}_{j}\left(t_{i}\right)\right)^{2}.
\]
We repeat this procedure 50 times to find the means and standard deviations
of the MSPE and MSEE for each method. Table \ref{tab:Comparison-between-Luo method and proposed}
reports the results at different values of $\rho_{c}$ and different
numbers of predictors, $p$. As reported, our proposed approach is
superior to the sigComp method in terms of MSPE and MSEE for $p=1,\,3$.
For example, when $p=3$ and $\rho_{c}=0.5$, the average MSPE and
MSEE of the sigComp are $0.2052$ and $0.1048$, which are much larger
than the corresponding values ($0.1282$ and $0.0286$) achieved by
MTOT. However, the performance of sigComp is comparable or even slightly
better than MTOT for $p=6$. For example, when $p=6$ and $\rho_{c}=0.5$,
the average MSPE is $0.1104$ for sigComp and $0.1121$ for MTOT.
The reason that sigComp performs slightly better for a larger $p$
is that it imposes sparsity when estimating the parameters, thus reducing
the chance of overfitting. Figure \ref{fig:Prediction-examples_GPsim}
illustrates prediction examples obtained by each method, along with
the true curve for different $p$. As illustrated both of the approaches
produce accurate predictions for $p=6$. 
\begin{table}
\caption{Comparison between the proposed method and the sigComp method proposed
by \citet{luo2017function}.\label{tab:Comparison-between-Luo method and proposed}}
\centering{}{\small{}}%
\begin{tabular}{|c|c|c|c|c|c|}
\hline 
 &  & \multicolumn{2}{c|}{\textit{\small{}sigComp}} & \multicolumn{2}{c|}{\textit{\small{}MTOT}}\tabularnewline
\hline 
{\small{}$p$} & {\small{}$\rho_{c}$} & {\small{}MSPE} & {\small{}MSEE} & {\small{}MSPE} & {\small{}MSEE}\tabularnewline
\hline 
\hline 
{\small{}$1$} & {\small{}0} & {\small{}0.7155 (}0.041{\small{}3)} & 0.6139 (0.0414) & \textbf{\small{}0.1039}{\small{} (0.00130)} & \textbf{\small{}0.0025}{\small{} (0.0001)}\tabularnewline
\hline 
\multirow{2}{*}{{\small{}$3$}} & {\small{}0} & 0.2052{\small{} (0.0089)} & 0.1048 (0.0087) & \textbf{\small{}0.1282}{\small{} (0.0064)} & \textbf{\small{}0.0286}{\small{} (0.0061)}\tabularnewline
\cline{2-6} 
 & {\small{}0.5} & {\small{}0.2086 (0.0105)} & 0.0934 (0.0106) & \textbf{\small{}0.1159}{\small{} (0.0034)} & \textbf{\small{}0.0140}{\small{} (0.0029)}\tabularnewline
\hline 
\multirow{2}{*}{{\small{}$6$}} & {\small{}0} & {\small{}0.1052 (0.0011)} & 0.0055 (0.0011) & \textbf{\small{}0.1051}{\small{} (0.0010)} & \textbf{\small{}0.0053}{\small{} (0.0003)}\tabularnewline
\cline{2-6} 
 & {\small{}0.5} & \textbf{0.1104} (0.0018) & \textbf{0.0106} (0.008) & {\small{}0.1121 (0.0035)} & {\small{}0.0141 (0.0043)}\tabularnewline
\hline 
\end{tabular}{\small \par}
\end{table}

\subsection{Simulation studies for image/ structured point-cloud or non-smooth
output}

\paragraph*{Case I\textendash Waveform surface simulation: \textmd{We simulate
waveform surfaces $\mathcal{Y}_{i}$ based on two input tensors, $\mathcal{X}_{1i}\in\mathbb{R}^{P_{11}\times P_{12}\times\cdots\times P_{1l_{1}}}$
and $\mathcal{X}_{2i}\in\mathbb{R}^{P_{21}\times P_{22}\times\cdots\times P_{2l_{2}}}\,$$\left(i=1,\cdots M\right)$,
where $M$ is the number of samples.}}

\begin{figure}
\begin{centering}
\subfloat[\label{fig:example of case 1}]{\includegraphics[bb=25bp 20bp 210bp 160bp,clip,width=0.45\columnwidth]{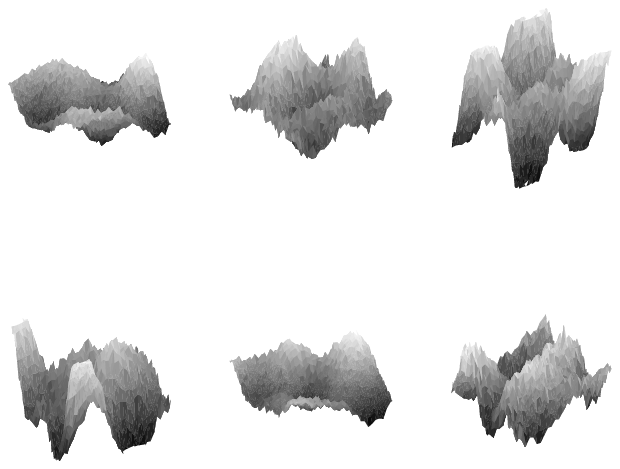}}\,~~~\subfloat[\label{fig:examplesOfCase2}]{\begin{centering}
\includegraphics[bb=30bp 30bp 200bp 150bp,clip,width=0.45\columnwidth]{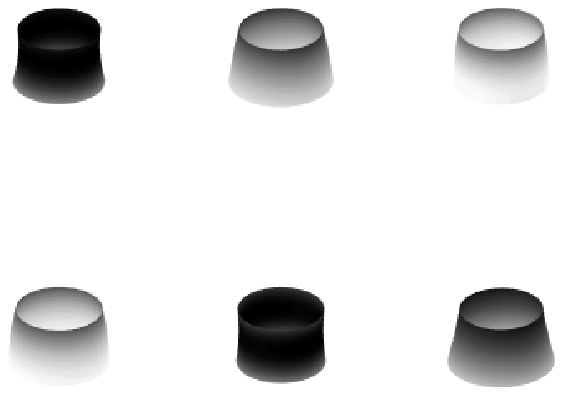}
\par\end{centering}
}
\par\end{centering}
\centering{}\subfloat[\label{fig:examplesOfCase3}]{\begin{centering}
\includegraphics[bb=15bp 0bp 210bp 170bp,clip,width=0.45\columnwidth]{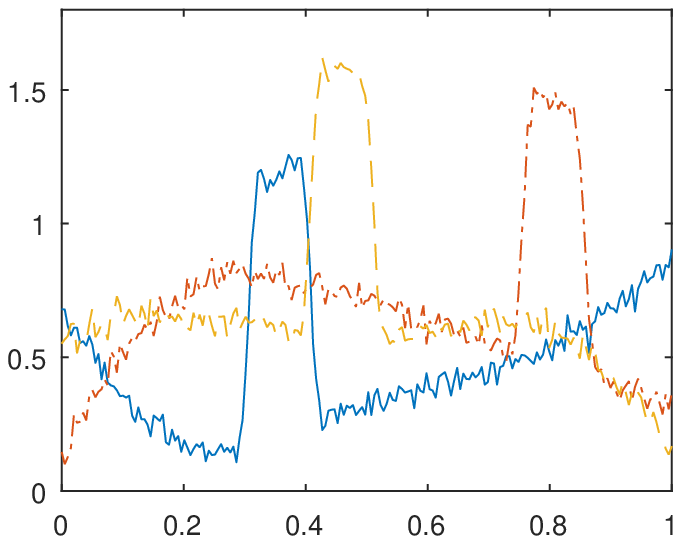}
\par\end{centering}

}\caption{Examples of generated output for the simulation study (a) case I-waveform
surface, (b) case II\textendash truncated cone, and (c) response function
with jump.}
\end{figure}

\paragraph*{\textmd{To generate the input tensors, we define $x_{kmj}=\frac{j}{P_{km}}\,\left(k=1,2;\,m=1,\cdots l_{k};\,j=1,\cdots P_{km}\right)$.
Then, we set $\boldsymbol{U}_{km}=\left[\boldsymbol{u}_{km1},\boldsymbol{u}_{km2},\cdots\boldsymbol{u}_{kmR_{k}}\right]\,\left(k=1,2;\,m=1,\cdots l_{k}\right)$,
where 
\[
\boldsymbol{u}_{kmt}=\protect\begin{cases}
\left[\cos\left(2\pi tx_{km1}\right),\cdots,\cos\left(2\pi tx_{kmP_{km}}\right)\right]^{\mathtt{T}} & \textrm{if \ensuremath{t} is odd}\protect\\
\left[\sin\left(2\pi tx_{km1}\right),\cdots,\sin\left(2\pi tx_{kmP_{km}}\right)\right]^{\mathtt{T}} & \textrm{if \ensuremath{t} is even}.
\protect\end{cases}
\]
Next, we randomly simulate elements of a core tensor $\mathcal{D}_{ki}$
from a standard normal distribution. Then, we generate an input sample
using the following model:
\[
\mathcal{X}_{ki}=\mathcal{D}_{ki}\times_{1}\boldsymbol{U}_{k1}\times_{2}\cdots\times_{l_{k}}\boldsymbol{U}_{kl_{k}}\,\left(k=1,2;\,i=1,\cdots,M\right).
\]
To generate a response tensor, we first simulate the elements of a
core tensor $\mathcal{C}_{k}$ from a standard normal distribution.
Moreover, we set $\boldsymbol{V}_{m}=\left[\boldsymbol{v}_{m1},\boldsymbol{v}_{m2},\cdots\boldsymbol{v}_{mR}\right]\,\left(m=1,\cdots,d\right)$,
where 
\[
\boldsymbol{v}_{mt}=\protect\begin{cases}
\left[\cos\left(2\pi ty_{m1}\right),\cdots,\cos\left(2\pi ty_{mQ_{m}}\right)\right]^{\mathtt{T}} & \textrm{if \ensuremath{t} is odd}\protect\\
\left[\sin\left(2\pi ty_{m1}\right),\cdots,\sin\left(2\pi ty_{mQ_{m}}\right)\right]^{\mathtt{T}} & \textrm{if \ensuremath{t} is even}
\protect\end{cases}
\]
 and $y_{mj}=\frac{j}{Q_{m}}$. Next, we define the parameter tensors
$\mathcal{B}_{k}$ using the following expansion: 
\[
\mathcal{B}_{k}=\mathcal{C}_{k}\times_{1}\boldsymbol{U}_{k1}\times_{2}\cdots\times_{l_{k}}\boldsymbol{U}_{kl_{k}}\times_{l_{k}+1}\boldsymbol{V}_{1}\times\cdots\times_{l_{k}+d}\boldsymbol{V}_{d}.
\]
Finally, we simulate a response tensor as
\[
\mathcal{Y}_{i}=\sum_{k=1}^{2}\mathcal{X}_{ki}*\mathcal{B}_{k}+\mathcal{E}_{i},
\]
where $\mathcal{E}_{i}$ is the error tensor whose elements are sampled
from a normal distribution $\mathcal{N}\left(0,\sigma^{2}\right)$.
For simulation purposes, we assume $\mathcal{X}_{1i}\in\mathbb{R}^{60}$,
$\mathcal{X}_{2i}\in\mathbb{R}^{50\times50}$, and $\mathcal{Y}_{i}=\mathbb{R}^{60\times40}$.
That is, we generate a response based on a profile and an image signal.
Furthermore, we set $R_{1}=2$, $R_{2}=3$, and $R=3$. This implies
that $\mathcal{C}_{1}\in\mathbb{R}^{2\times3\times3}$ and $\mathcal{C}_{2}\in\mathbb{R}^{3\times3\times3}$.
Figure \ref{fig:example of case 1} illustrates examples of generated
response surfaces. For this simulation study, we first generate a
set of $M=200$ data points. Then, we randomly divide the data into
a set of size $160$ for training and a set of size $40$ for testing.
We perform CV and train the model using the training set, then calculate
the SMSPE for the proposed method and benchmarks based on the testing
data. We repeat this procedure 50 times to capture the variance of
the SMSPE. In order to prepare data for the TOT approach, three steps
are performed: First, because the dimension of the curve inputs $\left(1\times60\right)$
and the image inputs $\left(50\times50\right)$ do not match, we randomly
select $50$ points out of 60 to reduce the curve dimension to 50.
Second, we replicate each curve 50 times to generate $50\times50$
images. Third, for each sample, we merge the image constructed from
the curve and the image input to construct a tensor of size $50\times50\times2$.
Combining all of the samples, we obtain an input tensor of size $M\times50\times50\times2$,
where $M$ is the sample size. }}

\paragraph*{Case II\textendash Truncated cone simulation: \textmd{We simulate
a truncated cone based on a set of scalars and simple profile data
in a 3D cylindrical coordinate system $\left(r,\phi,z\right)$, where
$\phi\in\left[0,2\pi\right]$ and $z\in\left[0,1\right]$. We first
generate an equidistant grid of $I_{1}\times I_{2}$ over the $\left(\phi,z\right)$
space by setting $\phi_{i}=\frac{2\pi i}{I_{1}}\,\left(i=1,\cdots,I_{1}\right)$
and $z_{j}=\frac{j}{I_{2}}\,\left(j=1,\cdots,I_{2}\right)$. Specifically,
we set $I_{1}=I_{2}=200$. Next, we simulate the truncated cone over
the grid by 
\begin{equation}
r\left(\phi,z\right)=\frac{r_{0}+z\tan\theta}{\sqrt{1-e^{2}\cos^{2}\phi}}+c\left(z^{2}-z\right)+\epsilon\left(\phi,z\right),\label{eq:truncated cone}
\end{equation}
where $r_{0}$ is the radii of the upper circle of the truncated cone,
$\theta$ is the angle of the cone, $e$ is the eccentricity of the
top and bottom surfaces, $c$ is the side curvatures of the truncated
cone, and $\epsilon\left(\phi,z\right)$ is process noise simulated
from $\mathcal{N}\left(0,\sigma^{2}\right)$. Figure \ref{fig:examplesOfCase2}
illustrates examples of generated truncated cones. We assume that
the parameters of the truncated cone are specific features obtained
from a scalar and three simple profile data. In particular, we assume
that the scalar predictor is $x_{1i}=r_{0i}$ and the profile predictors
are $x_{2i}\left(z\right)=z\tan\theta$, $x_{3i}\left(\phi\right)=e^{2}\cos^{2}\phi$,
and $x_{4i}\left(z\right)=c\left(z^{2}-z\right)$; $i=1,\cdots,M$.
That is, the inputs are one scalar and three profiles. We simulate
these profiles for training purposes by setting the parameters as
follows: We set $r_{0}\in\left\{ 1.1,1.3,1.5\right\} $, $\theta\in\left\{ 0,\frac{\pi}{8},\frac{\pi}{4}\right\} $,
$e\in\left\{ 0,0.3,0.5\right\} $, $c\in\left\{ -1,0,1\right\} $,
and consider a full factorial design to generate $81$ samples. That
is, for each combination of parameters (e.g., $\left\{ 1.1,\,0,\,0.3,\,-1\right\} $),
we generate a sample containing one scalar value and three profiles.
We represent each of the inputs by a matrix (a tensor of order 2)
to obtain four input matrices $X_{1}$, $X_{2}$, $X_{3}$, and $X_{4}$,
where $X_{1}\in\mathbb{R}^{81\times1}$ and $X_{i}\in\mathbb{R}^{81\times200}\,\left(i=2,3,4\right)$.
Finally, we generate the testing data by sampling the truncated cone
parameters as follows: We assume $r\sim U\left(1.1,1.5\right)$, $\theta\sim U\left(0,\frac{\pi}{4}\right)$,
$e\sim U\left(0,0.5\right)$, and $c\sim U\left(-1,1\right)$, where
$U\left(a,b\right)$ denotes a uniform distribution over the interval
$\left[a,b\right]$, and sample each parameter from its corresponding
distribution. In this simulation, we first train the model using the
generated training data. Next, we generate a set of 1000 testing data.
We predict the truncated cone based on the input values in the testing
data and calculate the SMSPE for each predicted cone. In order to
prepare the data for TOT, we first replicate the column of $X_{1}$
to generate a matrix of size $81\times200$, then merge this matrix
with the other three matrices to construct an input tensor of size
$81\times4\times200$. This tensor is used as an input in the TOT.}}

\paragraph*{Case III\textendash Curve response with jump simulation: \textmd{We
simulate a response function with jump using a group of B-spline bases.
Let $t_{i}=\frac{i}{I}$ with $i=0,1,\cdots,I$ and $I=200$ and let
$B_{1}\in\mathbb{R}^{I\times5}$ and $B_{2}\in\mathbb{R}^{I\times51}$
be two matrices of fourth-order B-spline bases obtained by one and
47 knots over $\left[0,1\right]$. We generate a response profile
by combining these two bases as follows:
\[
y_{j}\left(t_{i}\right)=B_{1}\left(t_{i}\right)x_{1j}+B_{2}\left(t_{i}\right)x_{2j}+e\left(t_{i}\right),
\]
where $B_{1}\left(t_{i}\right)$ and $B_{2}\left(t_{i}\right)$ are
basis evaluations at the point $t_{i}$, and $e\left(t_{i}\right)$
is a random error simulated from $\mathcal{N}\left(0,\sigma^{2}\right)$.
The input vector $x_{1j}$ is dense, and its elements are generated
from a uniform distribution over $\left[0,1\right]$. $x_{2j}$ is
a sparse vector with five consecutive elements equal to one and the
rest equal to zero. The location of five consecutive elements is selected
at random. Figure \ref{fig:examplesOfCase3} illustrates examples
of response functions. For this simulation study, we first generate
a set of $M=500$ data points, i.e., $\left\{ \left(y_{j},x_{1j},x_{2j}\right)\right\} _{j=1}^{M}$.
Then, we randomly divide the data into a set of size $400$ for training
and a set of size $100$ for testing. We perform CV and train the
model using the training data set. Next, we calculate the SMSPE for
the proposed method and the benchmark based on the testing data. We
repeat this procedure 50 times to capture the variance of the SMSPE.
}\protect \\
\textmd{\smallskip{}
}}

In each case, we compare the proposed method with benchmarks based
on the SMSPE calculated at different levels of noise $\sigma$. Tables
\ref{tab:Comparison-sim1-wavefrom-surface}, \ref{tab:Comparison-caseII-cone},
and \ref{tab:Comparision-CaseIII-jumpcurve} report the average and
standard deviation of SMSPE (or its logarithm), along with the average
running time of each algorithm for the simulation cases I, II, and
III, respectively. In Table \ref{tab:Comparison-caseII-cone}, we
report the average and standard deviation of the logarithm of the
SMSPE for better comparison of the values. Please notice that the
SMSPE is a standardized error and should not directly be compared
to the variance. In all cases, the MTOT has the smallest prediction
errors, reflecting the advantage of our method in terms of prediction.
Furthermore, with the increase in $\sigma$, all methods illustrate
a larger SMSPE in all cases. In the first case, the TOT illustrates
a prediction performance comparable to our method at a cost of a much
longer running time. For example, when $\sigma=0.2$, TOT requires
about 147.33 seconds to reach the SMSPE of 0.0170, obtained in 1.05
seconds by MTOT. The performance of both PCR and TOT are significantly
worse than MTOT in the second case. The inferior performance of TOT
is due to both its restriction on selecting the same rank for both
the input and output and the fact that the CP decomposition it uses
does not consider the correlation between multiple modes.

\begin{figure}
\centering{}\subfloat[\label{fig:example of cas3 params1}]{\includegraphics[width=0.5\columnwidth]{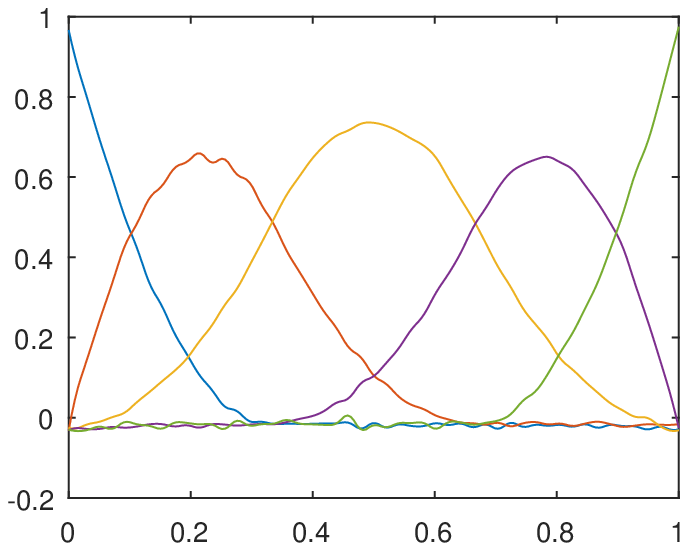}}\subfloat[\label{fig:examplesOfCase3 param2}]{\begin{centering}
\includegraphics[width=0.5\columnwidth]{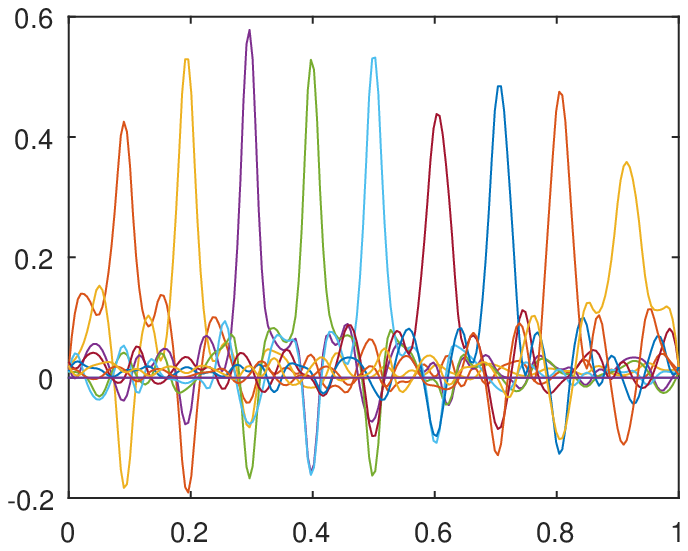}
\par\end{centering}
}\caption{Examples of the learned parameters in simulation case III when $\sigma=0$.}
\end{figure}

In the third case, the prediction performances of all three methods
are comparable, indicating that all three are capable of predicting
a functional output with discontinuity. However, our approach shows
slightly smaller prediction errors. Although the running time of the
PCR is significantly lower than the other two approaches, MTOT running
time is reasonable and within two-tenths of a second. The TOT shows
slightly larger prediction error in all cases with much longer running
time, making this approach less appealing. Recall that in this simulation,
we used B-spline bases as the coefficients of the input to generate
the output curve. Figure \ref{fig:example of cas3 params1} illustrates
the plot of the columns of the learned coefficient matrix that corresponds
to $B_{1}$. As can be seen, the learned bases are very similar to
B-spline bases used originally as the coefficients. Figure \ref{fig:examplesOfCase3 param2}
illustrates some of the columns of the learned parameters that correspond
to $B_{2}$. Unlike the first set of parameters, these parameters
are slightly different from the B-spline bases that are originally
used for data generation purposes. This is due to the identifiability
issue. Our approach imposes an orthogonality restriction that may
generate a set of parameters (when the identifiability issue exists)
different from the parameters from which\textbf{ }the data is originally
generated, but that can still produce accurate predictions in terms
of the mean value. 
\begin{table}
\caption{Comparison between the proposed method (MTOT) and the benchmarks in
case I with the waveform response. The TOT requires a much larger
running time to achieve the same level of prediction error as the
MTOT. \label{tab:Comparison-sim1-wavefrom-surface}}
\centering{}%
\begin{tabular}{|c|c|c|c|c|c|c|}
\hline 
 & \multicolumn{2}{c|}{\textit{\footnotesize{}PCR}} & \multicolumn{2}{c|}{\textit{\footnotesize{}TOT}} & \multicolumn{2}{c|}{\textit{\footnotesize{}MTOT}}\tabularnewline
\hline 
{\footnotesize{}$\sigma$} & {\footnotesize{}SMSPE} & {\footnotesize{}Time (sec)} & {\footnotesize{}SMSPE} & {\footnotesize{}Time (sec)} & {\footnotesize{}SMSPE} & {\footnotesize{}Time (sec)}\tabularnewline
\hline 
\hline 
{\footnotesize{}0.1} & {\footnotesize{}0.0057 (0.0015)} & {\footnotesize{}0.03 (0.00)} & {\footnotesize{}0.0046 (0.0011)} & {\footnotesize{}154.98 (17.94)} & \textbf{\footnotesize{}0.0044}{\footnotesize{} (0.0011)} & {\footnotesize{}1.05 (0.03)}\tabularnewline
\hline 
{\footnotesize{}0.2} & {\footnotesize{}0.0199 (0.0045)} & {\footnotesize{}0.04 (0.00)} & \textbf{\footnotesize{}0.0170}{\footnotesize{} (0.0039)} & {\footnotesize{}147.33 (2.47)} & \textbf{\footnotesize{}0.0170}{\footnotesize{} (0.0040)} & {\footnotesize{}1.05 (0.01)}\tabularnewline
\hline 
{\footnotesize{}0.3} & {\footnotesize{}0.0455 (0.0097)} & {\footnotesize{}0.04 (0.00)} & {\footnotesize{}0.0399 (0.0086)} & {\footnotesize{}149.03 (1.36)} & \textbf{\footnotesize{}0.0395}{\footnotesize{} (0.0086)} & {\footnotesize{}1.05 (0.02)}\tabularnewline
\hline 
{\footnotesize{}0.4} & {\footnotesize{}0.0773 (0.0233)} & {\footnotesize{}0.04 (0.00)} & {\footnotesize{}0.0678 (0.0135)} & {\footnotesize{}149.13 (0.96)} & \textbf{\footnotesize{}0.0673}{\footnotesize{} (0.0212)} & {\footnotesize{}1.05 (0.03)}\tabularnewline
\hline 
{\footnotesize{}0.5} & {\footnotesize{}0.1186 (0.0222)} & {\footnotesize{}0.04 (0.00)} & {\footnotesize{}0.1036 (0.0231)} & {\footnotesize{}146.17 (0.95)} & \textbf{\footnotesize{}0.1032}{\footnotesize{} (0.0203)} & {\footnotesize{}1.04 (0.01)}\tabularnewline
\hline 
{\footnotesize{}0.6} & {\footnotesize{}0.1670 (0.0327)} & {\footnotesize{}0.04 (0.00)} & {\footnotesize{}0.1456 (0.0309)} & {\footnotesize{}147.75 (1.81)} & \textbf{\footnotesize{}0.1454}{\footnotesize{} (0.0299)} & {\footnotesize{}1.03 (0.01)}\tabularnewline
\hline 
\end{tabular}
\end{table}
\begin{table}
\caption{Comparison between the proposed method and the benchmarks in case
II with truncated cone. Due to the difference between the input and
the output rank, the performance of the TOT is significantly worse
than the MTOT. The PCR is very fast in estimation, but the prediction
accuracy is not as appealing as the MTOT. \label{tab:Comparison-caseII-cone}}
\centering{}%
\begin{tabular}{|c|c|c|c|c|c|c|}
\hline 
 & \multicolumn{2}{c|}{\textit{\footnotesize{}PCR}} & \multicolumn{2}{c|}{\textit{\footnotesize{}TOT}} & \multicolumn{2}{c|}{\textit{\footnotesize{}MTOT}}\tabularnewline
\hline 
{\footnotesize{}$\sigma$} & {\footnotesize{}log(SMSPE)} & {\footnotesize{}Time (sec)} & {\footnotesize{}log(SMSPE)} & {\footnotesize{}Time (sec)} & {\footnotesize{}log(SMSPE)} & {\footnotesize{}Time (sec)}\tabularnewline
\hline 
\hline 
{\footnotesize{}0.01} & {\footnotesize{}-5.555 (0.986)} & {\footnotesize{}0.05 (0.00)} & {\footnotesize{}-5.249 (1.326)} & {\footnotesize{}23.58 (5.93)} & \textbf{\footnotesize{}-8.095}{\footnotesize{} (1.196)} & {\footnotesize{}3.82 (0.09)}\tabularnewline
\hline 
{\footnotesize{}0.02} & {\footnotesize{}-5.509 (0.937)} & {\footnotesize{}0.07 (0.00)} & {\footnotesize{}-5.197 (1.254)} & {\footnotesize{}27.94 (6.11)} & \textbf{\footnotesize{}-7.629}{\footnotesize{} (0.869)} & {\footnotesize{}3.92 (0.10)}\tabularnewline
\hline 
{\footnotesize{}0.03} & {\footnotesize{}-5.441 (0.879)} & {\footnotesize{}0.08 (0.00)} & {\footnotesize{}-5.127 (1.175)} & {\footnotesize{}29.11 (8.46)} & \textbf{\footnotesize{}-7.215}{\footnotesize{} (0.666)} & {\footnotesize{}3.93 (0.12)}\tabularnewline
\hline 
{\footnotesize{}0.04} & {\footnotesize{}-5.360 (0.819)} & {\footnotesize{}0.06 (0.00)} & {\footnotesize{}-5.048 (1.097)} & {\footnotesize{}33.61 (9.02)} & \textbf{\footnotesize{}-6.856}{\footnotesize{} (0.537)} & {\footnotesize{}3.95 (0.14)}\tabularnewline
\hline 
{\footnotesize{}0.05} & {\footnotesize{}-5.269 (0.762)} & {\footnotesize{}0.07 (0.00)} & {\footnotesize{}-4.963 (1.023)} & {\footnotesize{}34.29 (14.55)} & \textbf{\footnotesize{}-6.543}{\footnotesize{} (0.454)} & {\footnotesize{}3.99 (0.13)}\tabularnewline
\hline 
{\footnotesize{}0.06} & {\footnotesize{}-5.173 (0.710)} & {\footnotesize{}0.07 (0.00)} & {\footnotesize{}-4.875 (0.956)} & {\footnotesize{}37.43 (15.19)} & \textbf{\footnotesize{}-6.266}{\footnotesize{} (0.402)} & {\footnotesize{}3.95 (0.14)}\tabularnewline
\hline 
\end{tabular}
\end{table}
\begin{table}
\caption{Comparison between the proposed method and the benchmarks in case
III with a non-smooth response. All three methods produce reasonable
predictions. However, the MTOT performs slightly better than the benchmarks
within a reasonable running time. \label{tab:Comparision-CaseIII-jumpcurve}}
\centering{}%
\begin{tabular}{|c|c|c|c|c|c|c|}
\hline 
 & \multicolumn{2}{c|}{\textit{\footnotesize{}PCR}} & \multicolumn{2}{c|}{\textit{\footnotesize{}TOT}} & \multicolumn{2}{c|}{\textit{\footnotesize{}MTOT}}\tabularnewline
\hline 
{\footnotesize{}$\sigma$} & {\footnotesize{}SMSPE} & {\footnotesize{}Time (sec)} & {\footnotesize{}SMSPE} & {\footnotesize{}Time (sec)} & {\footnotesize{}SMSPE} & {\footnotesize{}Time (sec)}\tabularnewline
\hline 
\hline 
{\footnotesize{}0.1} & {\footnotesize{}0.0255 (0.0010)} & {\footnotesize{}0.02 (0.00)} & {\footnotesize{}0.0250 (0.0008)} & {\footnotesize{}17.67 (0.58)} & \textbf{\footnotesize{}0.0230}{\footnotesize{} (0.0007)} & {\footnotesize{}0.26 (0.01)}\tabularnewline
\hline 
{\footnotesize{}0.15} & {\footnotesize{}0.0503 (0.0018)} & {\footnotesize{}0.03 (0.00)} & {\footnotesize{}0.0510 (0.0018)} & {\footnotesize{}17.27 (0.20)} & \textbf{\footnotesize{}0.0496}{\footnotesize{} (0.0017)} & {\footnotesize{}0.25 (0.01)}\tabularnewline
\hline 
{\footnotesize{}0.2} & {\footnotesize{}0.0851 (0.0027)} & {\footnotesize{}0.04 (0.00)} & {\footnotesize{}0.0862 (0.0026)} & {\footnotesize{}17.25 (0.22)} & \textbf{\footnotesize{}0.0848}{\footnotesize{} (0.0028)} & {\footnotesize{}0.25 (0.01)}\tabularnewline
\hline 
{\footnotesize{}0.25} & {\footnotesize{}0.1260 (0.0043)} & {\footnotesize{}0.05 (0.00)} & {\footnotesize{}0.1271 (0.0042)} & {\footnotesize{}16.01 (2.32)} & \textbf{\footnotesize{}0.1255}{\footnotesize{} (0.0042)} & {\footnotesize{}0.27 (0.00)}\tabularnewline
\hline 
{\footnotesize{}0.3} & {\footnotesize{}0.1730 (0.0046)} & {\footnotesize{}0.05 (0.00)} & {\footnotesize{}0.1734 (0.0041)} & {\footnotesize{}16.33 (2.67)} & \textbf{\footnotesize{}0.1725}{\footnotesize{} (0.0046)} & {\footnotesize{}0.27 (0.01)}\tabularnewline
\hline 
{\footnotesize{}0.35} & {\footnotesize{}0.2305 (0.0051)} & {\footnotesize{}0.05 (0.00)} & {\footnotesize{}0.2333 (0.0066)} & {\footnotesize{}16.69 (0.83)} & \textbf{\footnotesize{}0.2230}{\footnotesize{} (0.0052)} & {\footnotesize{}0.27 (0.00)}\tabularnewline
\hline 
\end{tabular}
\end{table}

\section{Case Study \label{sec:Case-Study}}

\begin{figure}
\begin{centering}
\includegraphics[bb=100bp 70bp 650bp 480bp,clip,width=0.6\columnwidth]{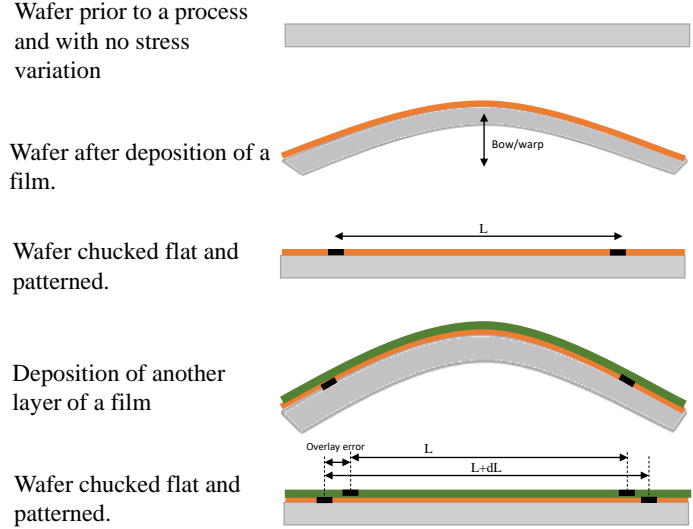}
\par\end{centering}
\caption{Process of a wafer, which causes shape variation and consequently
overlay error.\label{fig:Process-of-a-wafer-and overlayerror}}
\end{figure}

In semiconductor manufacturing, patterns are printed layer by layer
over a wafer in a sequence of deposition, etching, and lithographic
processes to manufacture transistors \citep{nishi2000handbook}. Many
of these processes induce stress variations across the wafer, distorting/changing
the wafer shape \citep{brunner2013characterization,turner2013role}.\textbf{
}Figure \ref{fig:Process-of-a-wafer-and overlayerror} illustrates
a simplified sequence of processes, causing the overlay error in the
patterned wafers. In the first step, a layer is deposited over the
wafer and exposed to rapid thermal annealing, causing a curvature
in the free-state wafer. The wafer is then chucked flat and patterned
in a lithographic process. Next, to generate a second layer pattern,
a new layer is deposited, changing the wafer shape. Finally, in the
lithography step, the flattened wafer is patterned. Because the wafer
is flattened, the first pattern distance increases, but the new pattern
is printed with the same distance $L$, generating a misalignment
between patterns. The overlay error caused by lower order distortions
can be corrected by most of the exposure tools. For this purpose,
the alignment positions of several targets are measured and used to
fit a linear overlay error model \citep{brunner2013characterization}:
\begin{equation}
\begin{cases}
\Delta x=T_{x}-\theta_{x}y+M_{x}x & \text{error in x coordinate}\\
\Delta y=T_{y}+\theta_{y}x+M_{y}y & \text{error in y coordinate,}
\end{cases}\label{eq:OverlayErrorModel}
\end{equation}
where $x$ and $y$ identify the position of the target point over
the wafer, $T_{x}$ and $T_{y}$ are transition errors, $\theta_{x}$
and $\theta_{y}$ relate to rotation error, and $M_{x}$ and $M_{y}$
are isotropic magnification errors pertaining to the wafer size change
or wafer expansion due to processing. The fitted model is then used
to correct the overlay errors. This model, however, can only correct
the overlay error induced by a uniform stress field and fails to compensate
for overlay errors caused by high-order distortions \citep{brunner2013characterization}.
Therefore, developing a model that can relate the overlay error to
higher order patterns in the wafer shape is essential for better overlay
correction.

In this case study, we use our proposed method to predict the overlay
error based on the wafer shape data. Such predictions can be fed forward
to the exposure tools to result in a better correction strategy. In
practice, the wafer shape is measured using a patterned wafer geometry
(PWG) tool, and the overlay error is measured using standard optical
methods \citep{brunner2013characterization}. Both the wafer shapes
and the overlay errors (in each coordinate, x or y) can be presented
as image data. In this case study, we follow the procedure and results
suggested and verified (through both experiments and finite element
{[}FE{]} analysis) by \citet{brunner2013characterization} to generate
surrogate data of overlay errors $\left(PIR\left(x,y\right)\right)$
based on the wafer shape prior to two lithography steps ($w_{1}\left(x,y\right)$,
$w_{2}\left(x,y\right)$). The data generation procedure is elaborated
in Appendix C. 

\begin{figure}[H]
\begin{centering}
\subfloat[\label{fig:PIR}]{\includegraphics[width=0.35\columnwidth]{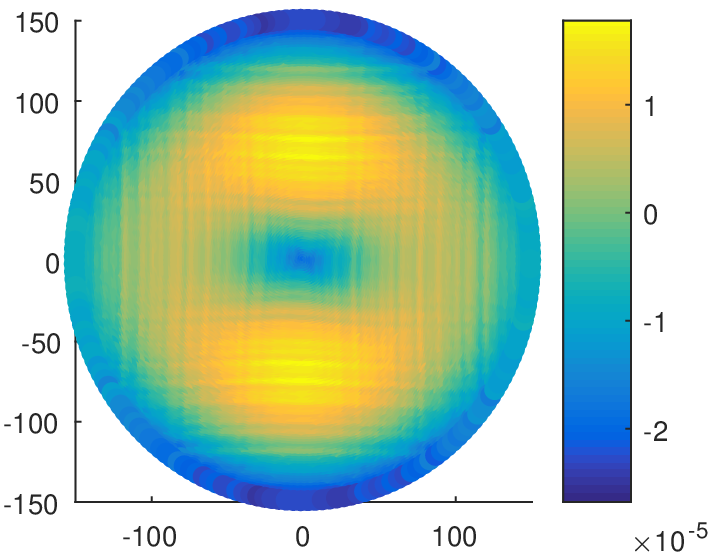}

}
\par\end{centering}
\begin{centering}
\subfloat[\label{fig:PIR_est}]{\includegraphics[width=0.35\columnwidth]{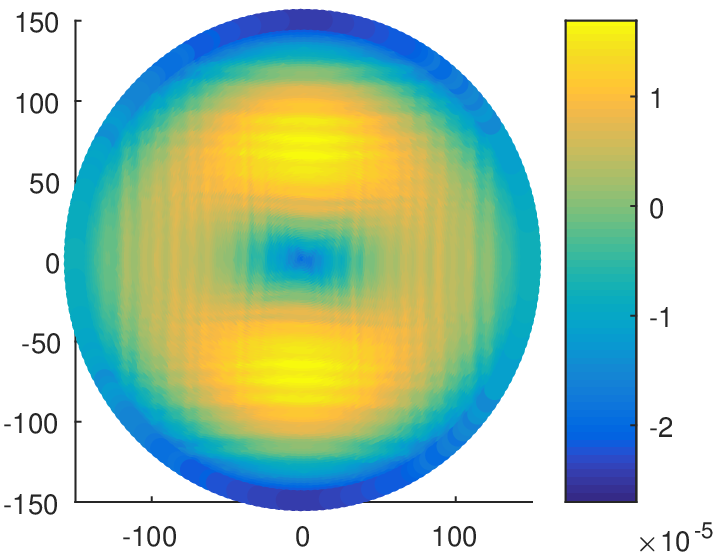}

}\subfloat[\label{fig:PIR-PIR_est}]{\begin{centering}
\includegraphics[width=0.35\columnwidth]{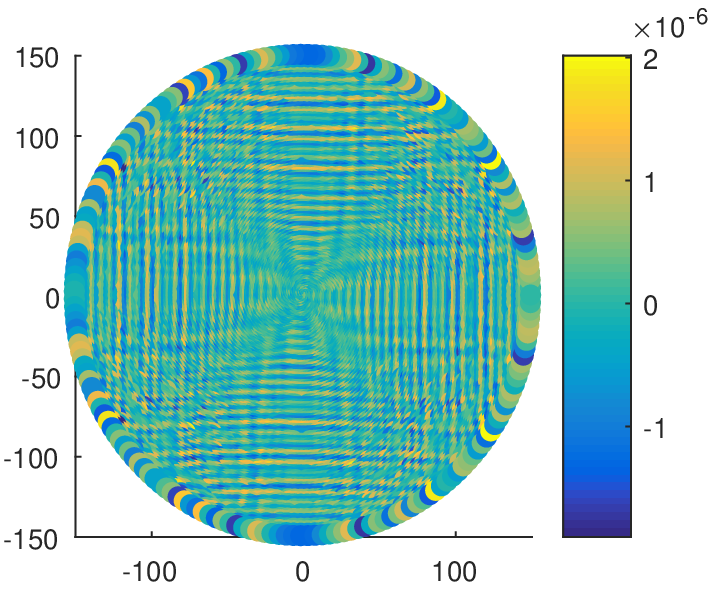}
\par\end{centering}
}
\par\end{centering}
\begin{centering}
\subfloat[\label{fig:PIR_est-1}]{\includegraphics[width=0.35\columnwidth]{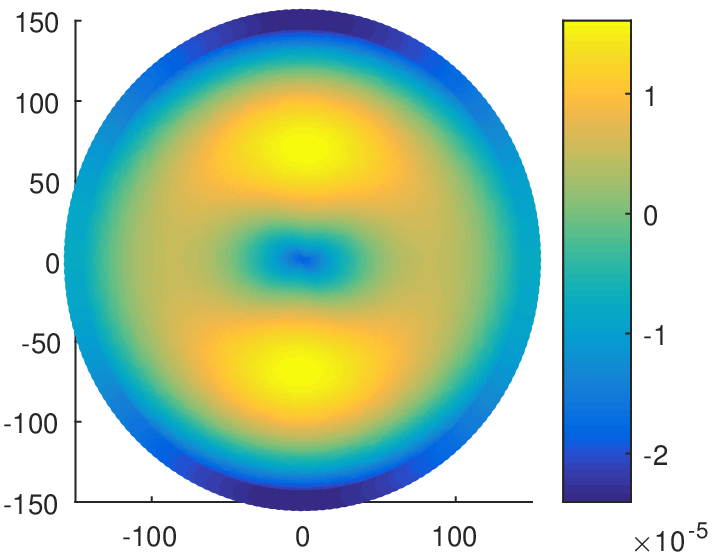}

}\subfloat[\label{fig:PIR-PIR_est-1}]{\begin{centering}
\includegraphics[width=0.35\columnwidth]{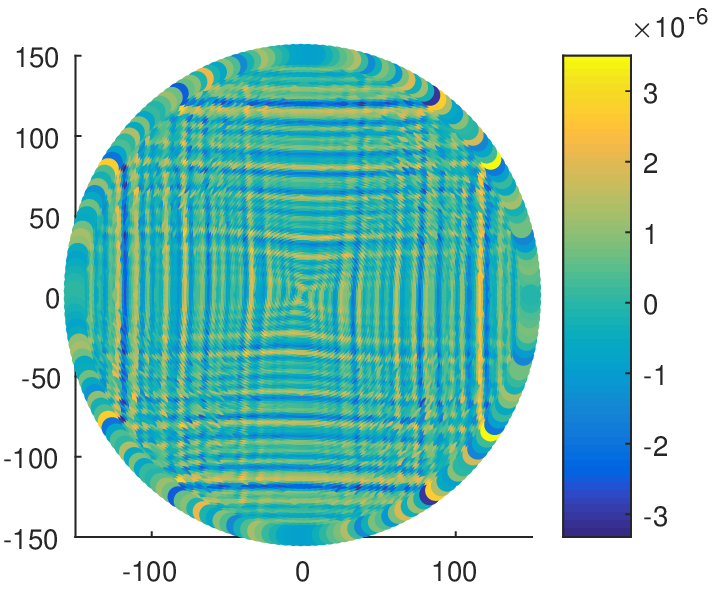}
\par\end{centering}
}
\par\end{centering}
\caption{Example of (a) the x coordinate overlay error, (b) prediction of MTOT,
(c) MTOT prediction error, (d) prediction of PCR, and (e) PCR prediction
error. \label{fig:Example-of-PIRandPIR_est} }
\end{figure}

\begin{figure}[H]
\begin{centering}
\includegraphics[bb=0bp 0bp 265bp 213bp,clip,width=0.7\columnwidth]{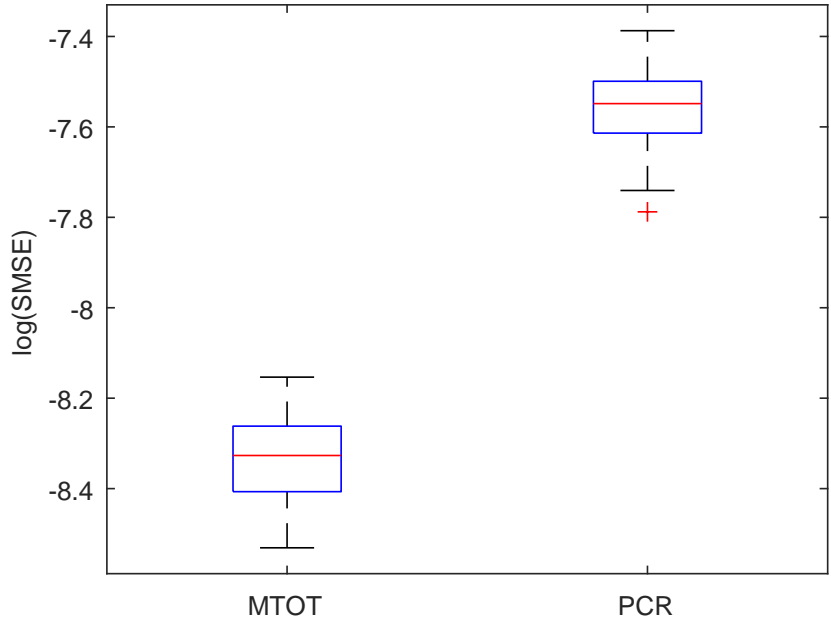}
\par\end{centering}
\caption{Logarithm of the prediction mean square error calculated for the test
data over 50 replications. The proposed method illustrates significantly
lower standard prediction error than the PCR approach. \label{fig:logarithm-of-prediction_error}}
\end{figure}

Based on the described procedure in Appendix C, we generate a set
of 500 training observations, i.e., wafer shapes and overlay errors,
$\left\{ \left(w_{1i}\left(x,y\right),w_{2i}\left(x,y\right),PIR_{i}\right)\right\} _{i=1}^{M=500}$,
and employ our proposed method to estimate the $PIR_{i}$ based on
$\left(w_{1i}\left(x,y\right),w_{2i}\left(x,y\right)\right)$. Because
in our simulated data $w_{1i}\left(x,y\right)$ remains fixed, we
consider $w_{i}\left(x,y\right)=w_{2i}\left(x,y\right)-w_{1i}\left(x,y\right)$
as the predictor. We also generate 100 observations as the test dataset.
The mean square prediction error obtained from the testing data is
used as the performance criterion. We repeat the simulations 50 times
and record the MSPE values. Because our proposed methodology assumes
that the shapes are observed over a grid, we transform the data to
the polar coordinate prior to modeling. In the polar space, each shape
is observed over a grid of $100\times200$ ($100$ in the radial direction
and $200$ in the angular direction, with overall 20,000 pixels).
Unfortunately, the TOT approach proposed by \citet{lock2017tensor}
failed to run with this size of images due to its high space complexity.
Therefore, we only compared our approach with PCR. Figure \ref{fig:Example-of-PIRandPIR_est}
illustrates an example of the original and predicted corrected overlay
error image, along with the prediction error. As illustrated, the
proposed method predicted the original surface more accurately, with
smaller errors across the wafer. Figure \ref{fig:logarithm-of-prediction_error}
illustrates the boxplots of the logarithm of the prediction mean square
error calculated over the 50 replications in contrast with the benchmark.
The results show that the proposed method is superior to the benchmark
in prediction of the image. As an example, the average of log(SMSE)
over the replications is -8.33 for the proposed method and -7.56 for
the PCR approach. 

\section{Conclusion\label{sec:Conclusion}}

This paper proposed a multiple tensor-on-tensor approach for modeling
processes with a heterogeneous set of input variables and an output
that can be measured by a scalar, curve, image, or point-cloud, etc.
The proposed method represents each of the inputs as well as the output
by tensors, and formulates a multiple linear regression model over
these tensors. In order to estimate the parameters, a least square
loss function is defined. In order to avoid overfitting, the proposed
method decomposes the regression parameters through a set of basis
matrices that spans the input and output spaces. Next, the basis matrices,
along with their expansion coefficients, are learned by minimizing
the loss function. The orthogonality condition is imposed over the
output bases to assure identifiability and interpretability. To solve
the minimization problem, first, a closed-form solution is derived
for both the bases and their coefficients. Second, the block coordinate
decent (BCD) approach combined with the ALS algorithm is applied.
The proposed approach is capable of combining different forms of inputs
(e.g., an image, a curve, and a scalar) to estimate an output (e.g.,
a scalar, a curve, or an image, etc.) as demonstrated in first three
simulation studies. For example, in the first and third simulation
studies, we combined the scalar and profile inputs to estimate a profile
and a point cloud, respectively; and in the second simulation study,
a profile and an image are integrated to predict an image.

In order to evaluate the performance of the proposed method, we conducted
four simulation studies and a case study. In our first simulation
study, we compared our proposed method with the function-on-function
approach proposed by \citet{luo2017function}. This simulation considered
scalar and curve inputs since the benchmark can only handle those
form of data. Next, we performed three other simulations to evaluate
the performance of the proposed method when the inputs or outputs
are images or point clouds. In these simulation studies, the proposed
approach was compared with principle component regression (PCR) and
tensor-on-tensor (TOT) regression, and showed superior performance
in terms of mean squared prediction error. We also evaluated our proposed
method using a set of surrogate data generated according to the manufacturing
process of semiconductors. We simulated the shape and overlay errors
for several wafers and applied the proposed method to estimate the
overlay errors based on the wafer shapes measured prior to the lithography
steps. Results showed that the proposed method performed significantly
better than the PCR in predicting the overlay errors.

As a future work, including penalties such as lasso for sparsity and
group lasso for variable selection and imposing roughness penalties
over the basis matrices may improve the prediction results and can
be further studied.

\section*{Appendix A: Proof of Proposition \ref{prop:MR-core tensor estimate}}

For simplicity, we assume only one input tensor exists. Then, we can
solve $\mathcal{C}$ by 

\begin{align*}
\textrm{argmin}_{\mathcal{C}}\|Y_{\left(1\right)}-X_{\left(1\right)}B\|_{F}^{2} & =\|Y_{\left(1\right)}-X_{\left(1\right)}(U_{l}\otimes U_{l-1}\otimes\cdots\otimes U_{1})C(V_{d}\otimes\cdots\otimes V_{1})^{T}\|_{F}^{2}\\
 & =\|vec(Y_{\left(1\right)})-vec(X_{\left(1\right)}(U_{l}\otimes U_{l-1}\otimes\cdots\otimes U_{1})C(V_{d}\otimes\cdots\otimes V_{1})^{T})\|_{2}^{2}\\
 & =\|vec(Y_{\left(1\right)})-vec(ZC(V_{d}\otimes\cdots\otimes V_{1})^{T})\|_{2}^{2}\\
 & =\|vec(Y_{\left(1\right)})-(V_{d}\otimes\cdots\otimes V_{1}\otimes Z)vec(C)\|_{2}^{2},
\end{align*}
 where $vec\left(X\right)$ stacks the columns of matrix $X$ on top
of each other. This is a simple least square regression that gives
a closed-form solution as in (\ref{eq:SolutionToRegression-1}) after
applying Proposition \ref{prop:proposition1} to convert Kronecker
products to tensor products.

\section*{Appendix B: Proof of Proposition \ref{prop:MR-basis estimate}}

Again assuming a single input tensor, let us define $\tilde{Y}=(V_{d}\otimes\cdots\otimes V_{1}\otimes Z)vec(C)$
in the tensor format as $\tilde{\mathcal{Y}}=\tilde{C}\times_{1}Z\times_{2}V_{1}\times\cdots\times_{d+1}V_{d}$,
which can be written as $\tilde{Y}_{(i)}=V_{i}\tilde{C}_{(i)}(V_{d}\otimes\cdots V_{i+1}\otimes V_{i-1}\otimes V_{1}\otimes Z)^{\mathtt{T}}$.
First note that 

\begin{align*}
\textrm{argmin}_{V_{i}}\|Y_{\left(1\right)}-X_{\left(1\right)}B\|_{F}^{2} & =\textrm{argmin}_{V_{i}}\|vec(Y_{\left(1\right)})-(V_{d}\otimes\cdots\otimes V_{1}\otimes Z)vec(\tilde{C})\|_{F}^{2}\\
 & =\textrm{argmin}_{V_{i}}\|Y_{(i)}-\tilde{Y}_{(i)}\|_{F}^{2}\\
 & =\textrm{argmin}_{V_{i}}\|Y_{(i)}-V_{i}\tilde{C}_{(i)}(V_{d}\otimes\cdots V_{i+1}\otimes V_{i-1}\otimes V_{1}\otimes Z)^{\mathtt{T}}\|_{F}^{2},
\end{align*}
with $A:=\tilde{C}_{(i)}(V_{d}\otimes\cdots V_{i+1}\otimes V_{i-1}\otimes V_{1}\otimes Z)^{\mathtt{T}}$.
Then, we want to solve 
\[
\textrm{argmin}_{V_{i}}\|Y_{(i)}-V_{i}A\|_{F}^{2}\,s.t.\,V_{i}^{\mathtt{T}}V_{i}=I.
\]
This is an orthogonal procrustes problem and is known to have solution
as is stated in Proposition \ref{prop:MR-basis estimate}.

\section*{Appendix C: Simulating the Overlay Error }

\citet{brunner2013characterization} introduced a measure based on
in-plane distortion (IPD) called predicted in-plane distortion residual
(PIR) to estimate and predict nonuniform-stress-induced overlay errors
based on wafer shape. For this purpose, they first illustrate that
the IPD is proportional to gradient of wafer shape $w\left(x,y\right)$,
i.e., 
\[
IPD\varpropto-\nabla w.
\]
Then, for two layers, say $i$ and $k$, to be patterned, they calculate
the IPD and subtract them to find the shape-slope difference, i.e.,
$SSD=IPD_{i}-IPD_{k}$. The shape-slope is then corrected based on
model (\ref{eq:OverlayErrorModel}) to find the shape-slope residual
(SSR). Finally, \citet{brunner2013characterization} calculated the
PIR as a factor of SSR. That is, 
\[
PIR=c\times SSR,
\]
where $c$ is a constant that depends on the wafer thickness. In their
study, they showed through four differently patterned engineer stress
monitor (ESM) wafers that the PIR is linearly correlated by the overlay
errors with high $R^{2}$ values (e.g., 92\%). To perform the experiment,
they first deposit a layer of silicon nitride film over a 300mm wafer
as a source of stress. This process changes the wafer shape and causes
the wafer to curve. The shape of the wafer is measured by a patterned
wafer geometry (PWG) tool designed for the metrology of 300mm wafers.
After the wafer is exposed by a designed pattern (four different patterns
considered in this study), it goes through an etching process that
relieves some part of the stress depending on the pattern density.
At this stage, and prior to next lithography step, the wafer shape
is again measured. After the second lithography step, the overlay
error is measured using standard optical methods. Using the measured
wafer shapes, they calculated the PIR and showed a high correlation
between the PIR and overlay error.

\begin{figure}[H]
\centering{}\includegraphics[bb=80bp 90bp 770bp 470bp,clip,width=0.7\columnwidth]{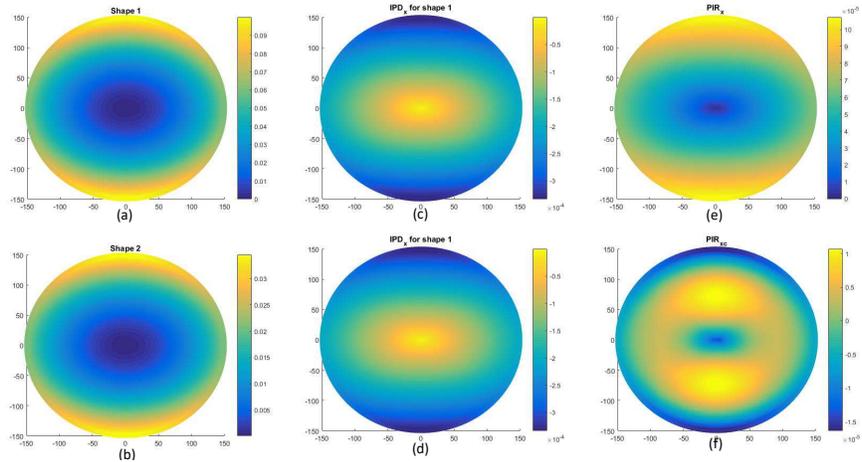}\caption{Illustration of wafer (a) shape prior to first step lithography, (b)
shape prior to second step lithography, (c) $IPD_{x}$ for the first
shape, (d) $IPD_{x}$ for the second shape, (e) PIR prior to correction,
and (f) PIR after correction for second order shapes.\label{fig:Illustration-of-wafer_shapes_IPD_IPR}}
\end{figure}

In our study, we first simulate the wafer shapes and then estimate
the overlay errors using the following procedure introduced by \citet{brunner2013characterization}.
Simulating a wafer shape requires knowledge of the different components
in wafer geometry. \citet{brunner2013characterization} and \citet{turner2013role}
consider several wafer shape features that span different ranges of
spatial wavelength $\left(\lambda\right)$. At the highest level is
the overall shape of the wafer represented by a bow (or warp) in the
range of tens of micrometers. Other shape variations are those that
are spanned by spatial wavelengths in the range of several meters
and waveheight in the micrometer range. Another component is the nanotopography
(NT) of the wafer, with $\lambda$ ranging from few millimeters to
20mm and the wave-height in nanometers. Finally, the roughness of
the wafer is defined as variations with $\lambda<0.2$mm. In this
study, we only consider the bow shape and the NT components when simulating
a wafer shape. The wafer shapes are simulated as follows: We first
assume that a thin layer is deposited over a wafer, which causes only
a bow shape geometry in the wafer (that is, we assume no wave patterns).
We simulate the bowed wafer geometry using $w_{1}\left(x,y\right)=\frac{b_{1}\left(0.5x^{2}+y^{2}\right)}{R^{2}}$,
where $b_{1}$ is the warp or bow size and is assumed to be $100\text{\ensuremath{\mu}m}$,
and $R$ is the wafer radius, which is assumed to be $150$mm. We
then assume that a lithography/etching process is performed and the
wafer shape changes in both bow and wavy patterns as follows:
\[
w_{2}\left(x,y\right)=\frac{b_{2}\left(0.5x^{2}+y^{2}\right)}{R^{2}}+\sum_{i=1}^{p}\frac{h_{i}}{2}\left(1+sin\left(\frac{2\pi x}{\lambda_{i}}\right)\right)+\sum_{i=1}^{p}\frac{h_{i}}{2}\left(1+cos\left(\frac{2\pi y}{\lambda_{i}}\right)\right),
\]
where $b_{2}$ is the bow size uniformly sampled from $30$ to $100\mu m$,
and $h_{i}$ and $\lambda_{i}$ are the waveheight and wavelength,
respectively. Moreover, $p$ is the number of waveforms assumed. For
each wafer (i.e. sample), we first randomly select $p$ from $U\left(2,10\right)$
and then select $p$ wavelength from $U\left(2,20\right)$ for NT
wavelength. Finally, we sample waveheight $h_{i}$ from $U\left(\frac{\lambda_{i}}{10^{7}},\frac{\lambda_{i}}{10^{6}}\right)$
to ensure that the large wavelength has large a waveheight and vice
versa. After simulating a wafer shape prior to two lithography steps,
we calculate the IPD and PIR according to the procedure described
previously. Note that we only calculate the values proportional to
the original values. Figure \ref{fig:Illustration-of-wafer_shapes_IPD_IPR}
illustrates an example of generated shapes, their associated IPDs
in the x coordinates, i.e., $-\frac{\partial w}{\partial x}$, and
the difference between the x coordinate IPDs prior to and after correction.
In order to correct the IPD values, we consider a second order model:
\[
\begin{cases}
\Delta IPD_{x}=k_{0}+k_{1}x+k_{2}y+k_{3}x^{2}+k_{4}y^{2}+k_{5}xy & \text{error in x coordinate}\\
\Delta IPD_{y}=k_{6}+k_{7}x+k_{8}y+k_{9}x^{2}+k_{10}y^{2}+k_{11}xy & \text{error in y coordinate},
\end{cases}
\]
which is fitted to the calculated values of $\Delta IPD_{x}=IPD_{x2}-IPD_{x1}$
and $\Delta IPD_{y}$. Then the fitted model is subtracted from the
$\Delta IPD_{x}$ and $\Delta IPD_{y}$ to find the corrected values.
The corrected values are associated with the PIR and the overlay error.

\section*{}

\bibliographystyle{authordate4}
\bibliography{../TOT_refs}

\end{document}